\documentclass[english,twocolumn, 10pt]{IEEEtran}
\usepackage[T1]{fontenc}
\usepackage[latin9]{inputenc}
\usepackage{babel}
\usepackage{array}
\usepackage{multirow}
\usepackage{amsmath}
\usepackage{amsthm}
\usepackage{amssymb}
\usepackage{stackrel}
\usepackage{graphicx}
\usepackage[unicode=true]
 {hyperref}

\makeatletter

\providecommand{\tabularnewline}{\\}

\theoremstyle{plain}
\newtheorem{thm}{\protect\theoremname}
\theoremstyle{plain}
\newtheorem{lem}[thm]{\protect\lemmaname}

\usepackage{etex}
\usepackage{url}
\usepackage{lmodern}
\usepackage[T1]{fontenc}
\usepackage{graphicx}
\usepackage{multicol}
\usepackage{multirow}
\usepackage{tikz, pgfplots, pgfkeys,amsmath, amssymb, psfrag}
\usepackage{cite}
\usepackage{algorithmic}
\usepackage{pifont}

\newcommand{\cmark}{\ding{51}}%
\newcommand{\xmark}{\ding{55}}%
\usetikzlibrary{patterns, positioning,arrows,calc}

\makeatother

\providecommand{\lemmaname}{Lemma}
\providecommand{\theoremname}{Theorem}

\begin{document}
\title{Side-Informed Steganography for JPEG Images by Modeling Decompressed
Images}
\author{Jan Butora and Patrick Bas\IEEEmembership{,~Senior Member,~IEEE}}
\maketitle
\begin{abstract}
Side-informed steganography has always been among the most secure
approaches in the field. However, a majority of existing methods for
JPEG images use the side information, here the rounding error, in
a heuristic way. For the first time, we show that the usefulness of
the rounding error comes from its covariance with the embedding changes.
Unfortunately, this covariance between continuous and discrete variables
is not analytically available. An estimate of the covariance is proposed,
which allows to model steganography as a change in the variance of
DCT coefficients. Since steganalysis today is best performed in the
spatial domain, we derive a likelihood ratio test to preserve a model
of a decompressed JPEG image. The proposed method then bounds the
power of this test by minimizing the Kullback-Leibler divergence between
the cover and stego distributions. We experimentally demonstrate in
two popular datasets that it achieves state-of-the-art performance
against deep learning detectors. Moreover, by considering a different
pixel variance estimator for images compressed with Quality Factor
100, even greater improvements are obtained.
\end{abstract}


\section{Introduction}

Steganography is a tool for covert communication, in which Alice and
Bob want to communicate secretly through a public channel. Their requirement
states that Eve, monitoring their public communication channel, should
not be aware of the secret communication. This requirement is achieved
by 'hiding' the communicated message in an ordinarily looking medium
- the cover object while preserving the semantic meaning of the cover.
Modifying a cover to carry the secret message then yields a stego
object. A typical cover medium is a digital image because of the vast
number of pixels and hard-to-model content. This difficulty of modeling
content allows Alice to create statistically undetectable modifications
and communicate secret messages of non-trivial size. While there are
dozens, maybe even hundreds, of possible ways to tackle this problem,
the state-of-the-art of hiding messages stems mainly from side-informed
steganography.

Side-informed steganography is a special case of covert communication
in which the steganographer (Alice) has access to the so-called side
information. The side information is usually defined as any additional
information about the cover image unavailable to the steganalyst (Eve).
Typically it results from some information-losing processing, such
as downsampling, JPEG compression, color conversion, etc., as long
as the last step of the processing is quantization. In these cases,
the side information consists of rounding errors because these information-losing
operations are always followed by quantizing the image coefficients
(whether DCT coefficients or pixels) to integers. Alice can then use
this information while creating her stego object to preserve better
the statistics of the original (precover) image before it was processed
and embedded with the secret message.

The most popular side information comes from JPEG compression~\cite{Fri05pq,But21PQ,Bor20ei,Den15wifs,Fri13spie,Gib20si2,Gib20si,Gib22TIFS,Gib22si}.
During JPEG compression, we change the representation of a given precover
image represented in the pixel domain with the Discrete Cosine Transformation
(DCT) (and consecutive quantization with a quantization matrix defined
by the JPEG quality factor), which is, in fact, an invertible operation.
However, this transformation produces non-integer DCT coefficients.
To perform the actual compression, the DCT coefficients are additionally
rounded to integers and only then saved to a JPEG file~\cite{Pen93}.
This loss of information makes the JPEG compression lossy, and the
signal that gets lost (the DCT rounding errors u ) is then used as
the side information. We want to note that many imaging devices today
round the quantized DCT coefficients to the nearest integer, but many
of them round toward zero instead~\cite{Far18wifs}.

One of the early uses of side information for steganography was employed
for double-compressed images~\cite{Fri05pq}. The side information,
in this case, would be created by recompressing a JPEG cover image,
and the embedding scheme would make changes only in coefficients with
the corresponding rounding errors close to $\pm1/2$. Such coefficients
are intuitively the most 'unstable' ones, meaning that even a tiny
perturbation can flip its value. In~\cite{But21PQ}, this idea was
extended into content-adaptive steganography by allowing changes only
in specific 'contributing' DCT modes producing many DCT coefficients
with rounding errors close to $\pm1/2$.

\begin{table*}
\caption{\label{tab:Comparison}Comparison of different side-informed schemes
with the proposed JEEP.}

\centering{}%
\begin{tabular}{c|cccccccc}
Algorithm & Stego model & Model domain & Emb. Domain & Modulation & SI use & RAW & Pipeline & |Alphabet|\tabularnewline
\hline 
SI-UNIWARD~\cite{Den15wifs} & N/A & N/A & spatial/JPEG & $(1-2|u|)$ & heuristic & \xmark & \xmark & $3$\tabularnewline
SI-MiPOD~\cite{Den17si} & Gauss. mixture & Pixel & spatial & $(1-2|u|)^{2}$ & model & \xmark & \xmark & $3$\tabularnewline
SI-JMiPOD~\cite{Gib22TIFS} & Gauss. mixture & DCT & JPEG & $(1-2|u|)^{4}$ & heuristic & \xmark & \xmark & $3$\tabularnewline
NS~\cite{NS19,NS20,NS16} & Gaussian & Pixel/DCT & spatial/JPEG & N/A & model & \cmark & \cmark & $\geq3$\tabularnewline
GE~\cite{Gib22si} & Gaussian & DCT & JPEG & N/A & model & \xmark & \cmark & $\geq3$\tabularnewline
JEEP & Gaussian & \multirow{1}{*}{Decompressed} & JPEG & Eq.~\eqref{eq:FM_si} & model & \xmark & \xmark & $3$\tabularnewline
\hline 
\end{tabular}
\end{table*}

\subsection{Side Information Usage in Literature}

\label{subsec:SI_literature}

With the development of content-adaptive steganography~\cite{Hol13eur,WOW12,Hill14,Sed15MiPOD,Guo15uerd,UED12,Shi14siued,Cog20color},
the use of side information has changed. The steganographic algorithms
output costs $\rho_{i}$ of changing the $i$-th cover element (pixel
or DCT coefficient) by $+1$ or $-1$, which are then used inside
Syndrome-Trellis Codes~\cite{Fil10stc} for near-optimal coding.
Having the embedding costs $\rho_{i}$ and the side information $u_{i}$,
it was proposed~\cite{Hol13ih} to heuristically modulate the cost
of changing an element 'towards the precover' by $1-2|u_{i}|$, where
$u_{i}$ is the rounding error introduced during the JPEG compression.
The embedding change in the opposite direction is prohibited, resulting
in a binary embedding scheme. This modulation technique was extended
for ternary embedding schemes by keeping the embedding cost in the
opposite direction intact~\cite{Den15wifs}. This creates asymmetry
in the ternary embedding, and the embedder ends with embedding costs
$\rho_{i}(\pm1)$ of changing the $i$-th element by $+1$ or $-1$:
\begin{eqnarray}
\rho_{i}(\mathrm{sign}(u_{i})) & = & \rho_{i}(1-2|u_{i}|),\label{eq:rho_plus_si}\\
\rho_{i}(-\mathrm{sign}(u_{i})) & = & \rho_{i}.\label{eq:rho_minus_si}
\end{eqnarray}

Based on an investigation of JPEG images compressed with the Trunc
quantizer~\cite{But20icassp,Far18wifs}, it was shown~\cite{But20ei}
that increasing the embedding cost in the opposite direction by a
factor of $1+2|u|$ provides additional improvements in security,
giving embedding costs:

\begin{eqnarray}
\rho_{i}(+1) & = & \rho_{i}(1-2u_{i}),\label{eq:rho_plus_si2}\\
\rho_{i}(-1) & = & \rho_{i}(1+2u_{i}).\label{eq:rho_minus_si2}
\end{eqnarray}

For binary MiPOD~\cite{Sed15MiPOD}, which uses steganographic Fisher
information instead of embedding costs, it was shown~\cite{Den17si}
that the Fisher information $I_{i}$ should be modulated by

\begin{eqnarray}
I_{i} & = & I_{i}(1-2|u_{i}|)^{2}.\label{eq:FI_si}
\end{eqnarray}

Even though \eqref{eq:FI_si} is the only side-informed modulation
derived from a model, the embedding was assumed binary for simplicity.

Recently, it was proposed for the JPEG variant of MiPOD~\cite{Gib22TIFS}
to modulate its Fisher information by

\begin{eqnarray}
I_{i} & = & I_{i}(1-2|u_{i}|)^{4}.\label{eq:FI_si2}
\end{eqnarray}

Yet, the embedding algorithm had to be used in a binary setting because
it was unclear what to do with the Fisher information in 'the opposite
direction.'

For JPEG side-informed steganography, it was proposed~\cite{Hol13eur}
to avoid embedding in the so-called 'rational' DCT modes $(0,0),(0,4),(4,0),(4,4)$,
whenever the rounding error is close to $\pm1/2$. One can easily
show that these modes can only produce rounding errors $\{\frac{k}{8},k\in\{-4,-3,\ldots,4\}\}$.
In effect, this causes a significant portion of the errors in these
modes to satisfy $|u_{i}|=1/2$, leading to embedding costs equal
to 0. Consequently, many of these coefficients would be changed, causing
easily exploitable artifacts in the embedding scheme. Boroumand et
al.~\cite{Bor20ei} made a further improvement to side-informed embedding
schemes by synchronizing the selection channel - making embedding
costs aware of already performed embedding changes during iterative
embedding on lattices. 

A different approach to side-informed steganography uses even more
information during embedding than just the precover. Having the out-of-camera
RAW image, Natural Steganography~\cite{NS16,NS17} (NS) points out
that processing of the RAW image creates natural dependencies between
pixels. These dependencies were further exploited for JPEG images~\cite{NS19,NS20}.
Recent work by Giboulot et al.~\cite{Gib20si} proposes to derive
a covariance matrix of the heteroscedastic noise naturally present
in images. This model is then extended to subsequent JPEG compression,
and the embedding algorithm aims to preserve the statistical model
of such noise. The method was extended in Gaussian Embedding~\cite{Gib20si2,Gib22si}
(GE), where the algorithm no longer requires the RAW image. Still,
other important information about the development has to be known,
namely ISO setting, camera model, and the processing pipeline. The
main drawback of these methods is that we need access to the RAW image
or the processing pipeline together with additional camera information.

\subsection{Comparison to Prior Art and Novelty}

We now point out the key differences and originality of the proposed
method w.r.t. prior art. In this work, for the first time (to the
best of our knowledge), we model both the decompressed JPEG image
and the effect of JPEG steganography in the spatial (pixel) domain.
Having these models, we propose a new steganographic method, JEEP
- JPEG Embedding preserving spatial Error Properties. We consider
the effect on the spatial domain because the state-of-the-art detectors
of JPEG steganography are Convolutional Neural Networks that operate
on decompressed images~\cite{You19ih,chubachi20wifs,yousfi20wifs,You21ih,But19ei,cogranne20wifs}.
Similarly to~\cite{Gib22si,NS19,NS20}, we show that the side information
emerges naturally from the proposed model. However, since we are modeling
the decompressed image, we do not need additional information on the
processing pipeline of the precover image, nor do we need to model
the discrete stego signal with a continuous distribution. 

Furthermore, we show that having the Fisher information $I_{i}$,
the right thing to do is to construct a Fisher information matrix
$\mathbf{I}_{i}=\begin{pmatrix}I_{i}^{+} & I_{i}^{\pm}\\
I_{i}^{\pm} & I_{i}^{-}
\end{pmatrix}$ with entries:

\begin{eqnarray}
I_{i}^{+} & = & I_{i}(1-2u_{i})^{4},\nonumber \\
I_{i}^{-} & = & I_{i}(1+2u_{i})^{4},\nonumber \\
I_{i}^{\pm} & = & I_{i}(1-2u_{i})^{2}(1+2u_{i})^{2}.\label{eq:FM_si}
\end{eqnarray}

We can notice that this very much resembles~\eqref{eq:FI_si2}; however,
in~\cite{Gib22TIFS}, the modulation is introduced heuristically
and does not allow for ternary embedding. Additionally, in~\cite{Gib22TIFS},
the image is modeled in the DCT domain as a mixture of distributions
since the stego signal cannot be assumed normally distributed. In
this work, we model the image in the spatial domain, which allows
us to consider steganography as a change of variance. Further comparison
is given in Table~\ref{tab:Comparison}. 

The main contribution of this paper can be summarized by the following:
\begin{itemize}
\item JPEG steganography is modeled in the pixel domain, where steganalysis
performs the best.
\item The side information cannot improve security against an omniscient
attacker. Hence it is crucial to constrain the attacker's knowledge.
\item We design a model-based side-informed scheme without additional knowledge
about the precover. This scheme has the side information naturally
present in the variance of stego distribution.
\item The proposed method is on par with or outperforms current state-of-the-art
steganography across all quality factors and payloads with Deep Learning
and Feature-based steganalyzers. 
\item By considering a different pixel variance estimator for high-quality
JPEG images, the proposed algorithm is also highly robust against
the Reverse JPEG Compatibility Attack~\cite{Cog20rjca,But20rjca}.
\end{itemize}

\subsection{Organization of the paper}

The rest of the paper is organized as follows: In the next section,
we introduce the proposed cover and stego model of a decompressed
JPEG image. Section~\ref{sec:Minimizing_power} presents the proposed
embedding scheme, JEEP, designed to preserve the cover model. In Section~\ref{sec:Dataset-and-detectors},
we introduce the datasets and detectors used for empirical evaluation
of its security. Numerical results are shown in Section~\ref{sec:Results},
and the paper is concluded in Section~\ref{sec:Conclusions}.

\section{Proposed Image Model}

\label{sec:Proposed-Image-Model}

This section first introduces the notation we keep using throughout
the paper and the basic mechanisms of JPEG compression. Then we present
a statistical model for decompressed cover and stego images in the
pixel domain. Finally, we will show that the side information available
to Alice is naturally contained in the variance of the stego pixels.
To derive our cover and stego models, we use several simplifying assumptions
introduced at the beginning of their respective sections. The validity
of these assumptions is then discussed at the end of their sections.

\subsection{JPEG Compression and Notation}

\label{sec:Preliminaries}Boldface symbols are reserved for matrices
and vectors. Uniform distribution on the interval $[a,b]$ is denoted
$\mathcal{U}[a,b]$ while $\mathcal{N}(\mu,\sigma^{2})$ is used for
the Gaussian distribution with mean $\mu$ and variance $\sigma^{2}$.
The operation of rounding $x$ to an integer is the square bracket
$[x]$. The sets of all integers and real numbers are denoted $\mathbb{Z}$
and $\mathbb{R}$. Expectation and variance of a random variable $X$
are denoted as $\mathbb{E}[X]$ and $\mathrm{Var}(X)$. Covariance
between two random variables $X$ and $Y$ is denoted as $\mathrm{Cov}(X,Y)$.
The original uncompressed 8-bit grayscale image with $N$ pixels is
denoted $\mathbf{x}\in\{0,\ldots,255\}^{N}$. For simplicity, we assume
that the width and height of the image are multiples of 8. 

Constraining $\mathbf{x}=(x_{ij})$ into one specific $8\times8$
block, we use indices $0\leq i,j\leq7$ to index elements from this
pixel block. Conversely, indices $0\leq k,l\leq7$ are strictly used
to index elements in the DCT domain. During JPEG compression, the
DCT coefficients before quantization, $d_{kl}\in\mathbb{R}$, are
obtained using the 2D-DCT transformation $d_{kl}=\mathrm{DCT}_{kl}(\mathbf{x})=\sum_{i,j=0}^{7}f_{kl}^{ij}x_{ij}$,
where

\begin{equation}
f_{kl}^{ij}=\frac{\omega_{k}\omega_{l}}{4}\cos\frac{\pi k(2i+1)}{16}\cos\frac{\pi l(2j+1)}{16},\label{eq:dct_cosines}
\end{equation}

$\omega_{0}=1/\sqrt{2},\omega_{k}=1$ for $0<k\leq7$ are the discrete
cosines. Before applying the DCT, each pixel is adjusted by subtracting
128 from it during JPEG compression, a step we omit here for simplicity.

The quantized DCT coefficients are $c_{kl}=[d_{kl}/q_{kl}]$, $c_{kl}\in\{-1024,\ldots,1020\}$,
where $q_{kl}$ are the quantization steps in a luminance quantization
table, provided in the header of the JPEG file.

Sometimes, it will be helpful to represent an $8\times8$ block of
elements with a vector of 64 dimensions instead. To this end, we denote
$\mathbf{D}\in\mathbb{R}^{64\times64}$ the matrix representing the
DCT transformation, and $\mathbf{c}\in\mathbb{R}^{64}$ a vector containing
the DCT coefficients. Additionally, we denote $\mathbf{Q}\in\mathbb{R}^{64\times64}$
the quantization matrix, containing the quantization steps on the
diagonal.

Compression and decompression can then be written as 

\begin{eqnarray}
\mathbf{c} & = & \left[\mathbf{Q}^{-1}\mathbf{D}\mathbf{x}\right],\label{eq:compression}\\
\mathbf{y} & = & \mathbf{D}^{T}\mathbf{Qc},\label{eq:decompression}
\end{eqnarray}

where $\mathbf{y}\neq\mathbf{x}$ represents the decompressed image.

To prevent confusion, we will be denoting $\Sigma_{ij}$ and $\Sigma_{kl}$
the diagonal covariance matrices in the spatial and DCT domains, respectively\footnote{Notice the slight abuse of notation since $ij$ an $kl$ are not used
as indices.}. The elements on their diagonals will be denoted as $\sigma_{ij}^{2}$
and $\sigma_{kl}^{2}$, $0\leq i,j,k,l\leq7$. To propagate pixel
variances into the DCT domain, we compute 

\begin{equation}
\Sigma_{kl}=\mathbf{D}\Sigma_{ij}\mathbf{D}^{T},\label{eq:spatial_to_dct_var}
\end{equation}

and similarly, to compute spatial variances from the DCT covariance
matrix

\begin{equation}
\Sigma_{ij}=\mathbf{D}^{T}\Sigma_{kl}\mathbf{D}.\label{eq:dct_to_spatial_var}
\end{equation}

\subsection{Cover Model}

\label{subsec:cover_model}

Recall, in~\cite{But20rjca}, the DCT rounding errors $u_{kl}$ are
modeled with a uniform distribution $\mathcal{U}(-1/2,1/2)$, and
thus spatial domain rounding errors $y_{ij}-[y_{ij}]$ are modeled
with Wrapped Gaussian distribution $\mathcal{N}_{W}(0,s_{ij})$, where
$s_{ij}=\sum_{k,l=0}^{7}\left(f_{kl}^{ij}\right)^{2}q_{kl}^{2}\mathrm{Var}(u_{kl})$.
That initially motivated this work because we realized that to preserve
the Wrapped Gaussian distribution of the model (see~\cite{But20rjca}
for more details), the steganographer (Alice) can maintain the underlying
Gaussian, which is computed as $y_{ij}-x_{ij}\sim\mathcal{N}(0,s_{ij})$.
Moreover, since we want to create a secure embedding scheme, we do
not have to limit ourselves only to quality factors (QFs) 99 and 100,
as was the case for Reverse JPEG Compatibility Attack (RJCA)~\cite{But20rjca,Cog20rjca}.

In this work, instead of modeling the decompression error, we will
model the decompressed pixels with full knowledge of the DCT rounding
errors. In turn, by preserving the pixel model, we shall also be preserving
the rounding error model as a consequence. But first, let us mention
several simplifying assumptions for our image model, which we will
comment upon at the end of this section:
\begin{itemize}
\item (C1) Uncompressed pixels are independent random variables $X_{ij}$
with $\mathbb{E}[X_{ij}]=x_{ij}$ and $\mathrm{Var}(X_{ij})=\sigma_{ij}^{2}$.
\item (C2) DCT rounding errors are independent random variable $U_{kl}$
with $\mathbb{E}[U_{kl}]=u_{kl}$ and $\mathrm{Var}(U_{kl})=0$.
\item (C3) Uncompressed pixels are independent of DCT rounding errors.
\end{itemize}
Note that we have not made any assumptions on the distributions of
the random variables so far because we consider the Central Limit
Theorem causing the decompressed pixels to be Normally distributed.
We can express the decompressed pixels as

\begin{eqnarray}
\mathbf{y} & = & \mathbf{D}^{T}\mathbf{Q}\mathbf{c}\nonumber \\
 & = & \mathbf{D}^{T}\mathbf{d}-\mathbf{D}^{T}\mathbf{Q}\mathbf{u}\nonumber \\
 & = & \mathbf{x}-\mathbf{D}^{T}\mathbf{Q}\mathbf{u}.\label{eq:cover_pixels}
\end{eqnarray}

We can therefore model them with 

\begin{equation}
\mathbf{y}\sim\mathcal{N}(\mathbf{x}-\mathbf{D}^{T}\mathbf{Qu},\Sigma_{ij}),\label{eq:cover_distribution}
\end{equation}

where $\Sigma_{ij}$ is the diagonal covariance matrix of the precover.

We used several assumptions (C1-C3) to derive meaningful cover image
models, and we would like to address their reasoning: 
\begin{itemize}
\item (C1) The independence assumption is made to simplify the model. Alternatively,
we can consider the covariance between pixels as part of a modeling
error when estimating their variances since they are not available
in practice.
\item (C2) We investigated intra-block and inter-block correlations of the
DCT rounding errors, but we did not find any evidence of correlation.
\item (C3) Similarly, as in (C1-C2), we did not find any evidence of correlation.
Alternatively, they could also be considered as part of a modeling
error.
\end{itemize}

\subsection{Stego Model}

\label{subsec:Stego-model}

To further simplify the situation, we make additional assumptions
on the embedding changes, which we will discuss at the end of the
section:
\begin{itemize}
\item (S1) Embedding changes are mutually independent.
\item (S2) Embedding changes are correlated with DCT errors.
\end{itemize}
We can model embedding changes $\eta_{kl}\in\{-1,0,1\}$ as random
variables with $P(\eta_{kl}=\pm1)=\beta_{kl}^{\pm}$ and $P(\eta_{kl}=0)=1-\beta_{kl}^{+}-\beta_{kl}^{-}$,
where $\beta_{kl}^{\pm}$ is the change rate (of change by +1 or -1).
This implies 
\begin{eqnarray}
\mathbb{E}[\eta_{kl}] & = & \beta_{kl}^{+}-\beta_{kl}^{-},\label{eq:embedding_mean}\\
\mathrm{Var}(\eta_{kl}) & = & \beta_{kl}^{+}+\beta_{kl}^{-}-(\beta_{kl}^{+}-\beta_{kl}^{-})^{2}.\label{eq:embedding_variance}
\end{eqnarray}

The stego image pixel values $\mathbf{z}$ can be expressed as
\begin{eqnarray}
\mathbf{z} & = & \mathbf{D}^{T}\mathbf{Q}(\mathbf{c}+\boldsymbol{\eta})\nonumber \\
 & = & \mathbf{D}^{T}\mathbf{d}+\mathbf{D}^{T}\mathbf{Q}(\boldsymbol{\eta}-\mathbf{u})\nonumber \\
 & = & \mathbf{x}+\mathbf{D}^{T}\mathbf{Q}\boldsymbol{\eta}-\mathbf{D}^{T}\mathbf{Qu}.\label{eq:stego_pixels}
\end{eqnarray}

Assuming the Central Limit Theorem again, we can now model the decompressed
stego pixels as Gaussian variables with mean

\begin{eqnarray}
\mathbb{E}[\mathbf{z}] & = & \mathbf{x}+\mathbf{D}^{T}\mathbf{Q}(\boldsymbol{\beta}^{+}-\boldsymbol{\beta}^{-}-\mathbf{u}),\label{eq:stego_mean}
\end{eqnarray}

and variance

\begin{eqnarray}
\overline{\Sigma}_{ij} & = & \Sigma_{ij}+\mathbf{D}^{T}\mathbf{Q}^{2}(\mathbf{E}-2\mathbf{C})\mathbf{D},\label{eq:stego_variance}
\end{eqnarray}

where $\mathbf{E}$ is the diagonal covariance matrix of the embedding
changes and $\mathbf{C}$ is the diagonal covariance matrix between
embedding changes and the DCT errors. 

We will see later in Section~\ref{subsec:Attacker_effect} that Eve's
knowledge of the expectation of embedding changes will severely affect
Alice's embedding strategy. 

\subsubsection{Variance}

One of the most significant contributions of this work is exploiting
the dependence between embedding changes and the side information.
Unfortunately, the covariance between the embedding changes and the
rounding errors $\mathrm{Cov}(\eta,U)=\mathbb{E}[\eta\cdot U]-\mathbb{E}[\eta]\cdot\mathbb{E}[U]$
is impossible to compute because we do not know how to calculate the
joint expectation. Instead, we will use several assumptions that will
allow us to approximate the covariance. For the ease of the following
derivations, we assume
\begin{itemize}
\item (S3) $\beta^{+}\geq2\beta^{-}$ (or $\beta^{-}\geq2\beta^{+}$ ).
\item (S4) $\mathbb{E}[\eta\cdot U]\geq0$.
\end{itemize}
It follows from Lemma~\ref{lem:lemma1} that $3\mathbb{E}[\eta]\cdot\mathbb{E}[U]\geq2\mathbb{E}[\eta^{2}]\cdot\mathbb{E}[U^{2}].$
We want to point out that in our experiments, we have observed this
inequality violated only when both change rates are either close to
1/3 or 0. Combining Lemma~\ref{lem:lemma1} with (S4), we will approximate
the joint probability as

\begin{eqnarray}
\mathbb{E}[\eta\cdot U] & = & 3\mathbb{E}[\eta]\cdot\mathbb{E}[U]-2\mathbb{E}[\eta^{2}]\cdot\mathbb{E}[U^{2}].\label{eq:joint_prob}
\end{eqnarray}

The covariance can then be expressed as

\begin{eqnarray}
\mathrm{Cov}(\eta,U) & = & 2(\beta^{+}-\beta^{-})u-2(\beta^{+}+\beta^{-})u^{2}.\label{eq:covariance}
\end{eqnarray}

The approximation~\eqref{eq:joint_prob} is reasonable because, with
a bigger expectation of embedding change, we get a bigger correlation
with the side information. Moreover, we would obtain a slight negative
correlation for symmetric change rates, which is logical because the
expectation of embedding change would be zero irrespectively of the
rounding error magnitude. Finally, for rounding errors close to zero,
the correlation would also be very close to zero. 

Plugging \eqref{eq:covariance} into \eqref{eq:stego_variance} reveals
that we can write the covariance matrix of the decompressed stego
image as

\begin{eqnarray}
\overline{\Sigma}_{ij} & = & \Sigma_{ij}+\mathbf{D}^{T}\mathbf{Q}^{2}\left(\boldsymbol{\beta}^{+}(1-2\mathbf{u})\right.\nonumber \\
 &  & \left.+\boldsymbol{\beta}^{-}(1+2\mathbf{u})-(\boldsymbol{\beta}^{+}-\boldsymbol{\beta}^{-})^{2}\right)\mathbf{D}.\label{eq:stego_variance2}
\end{eqnarray}

We can notice that this already resembles the usage of the side information
proposed in~\eqref{eq:rho_plus_si2},\eqref{eq:rho_minus_si2}. 

As with the cover model, we will now elaborate on the stego assumptions:
\begin{itemize}
\item (S1) In theory, this assumption is wrong because, during decompression,
the change rates get mixed together. We assumed independence across
change rates; otherwise, the optimization problem in~\eqref{eq:minimization_KLD}
involves computing an extremely numerically unstable hessian matrix
of size $128\times128$ for every $8\times8$ block. This turned out
to be relatively computationally heavy. However, it is something we
will consider solving in the future.
\item (S2) Without this correlation, we cannot fully utilize the side information
during embedding.
\item (S3) In other cases, the two change rates can be assumed to be relatively
close to each other. Thus it is reasonable to think that the side
information does not bring additional security to the model. We can
imagine those are the cases in which embedding does not change the
underlying model, such as in extremely noisy areas of the image. For
ease of the following derivations, we thus consider the other cases
negligible. 
\item (S4) Without this assumption, the variance in~\eqref{eq:stego_variance}
rises too rapidly with increasing change rates, and we cannot efficiently
preserve the cover distribution~\eqref{eq:cover_distribution}~\cite{But22SVP}.
\end{itemize}

\section{Minimizing Power of the Most Powerful Detector}

\label{sec:Minimizing_power}

In this section, we will derive Alice's embedding strategy to minimize
the power of the most powerful detector, which will turn out to be
the Likelihood Ratio Test (LRT). We consider two types of attackers
(steganalysts) - omniscient and realistic (constrained) Eve. We will
see that the two embedding strategies Alice can employ differ fundamentally
depending on Eve's capabilities. We call the resulting embedding algorithm
JEEP - JPEG Embedding preserving spatial Error Properties.

First, somewhat unrealistically, we will assume omniscient Eve, who
has complete knowledge of the system, including the precover and the
side information. This assumption is rather silly because if Eve knows
the precover, she can compute from it the cover image and decide whether
the image under investigation is a cover or not. We do this in order
to derive a general form of the LRT, which we will restrict afterward
with more realistic assumptions on Eve's abilities.

\subsection{Likelihood Ratio Test}

To make the test easier to follow, we will investigate the behavior
of the pixel residual $\mathbf{e}=\mathbf{y}-\mathbf{x}+\mathbf{D}^{T}\mathbf{Qu}$.
In this case, Eve's goal is to decide between the following two hypotheses:

\begin{eqnarray}
\mathcal{H}_{0} & : & \mathbf{e}\sim\mathcal{N}(0,\Sigma_{ij}),\label{eq:null_hypothesis_pixels}\\
\mathcal{H}_{1} & : & \mathbf{e}\sim\mathcal{N}(\boldsymbol{\mu},\overline{\Sigma}_{ij}),\label{eq:alt_hypothesis_pixels}
\end{eqnarray}

where $\boldsymbol{\mu}=\mathbf{D}^{T}\mathbf{Q}(\boldsymbol{\beta}^{+}-\boldsymbol{\beta}^{-})$
is the expectation of embedding change after decompression. Following
the Neyman-Pearson Lemma~\cite{lehmann05}, we will be interested
in the Likelihood Ratio Test (LRT), which can be used to minimize
detection power for a prescribed false alarm:

\begin{equation}
\Lambda(\mathbf{e})=\text{\ensuremath{\sum_{i=1}^{N}\Lambda(e_{i})}}=\sum_{i=1}^{N}\log\frac{p(e_{i},\mu_{i},\overline{\sigma}_{i}^{2})}{p(e_{i},0,\sigma_{i}^{2})}\stackrel[\mathcal{H}_{0}]{\mathcal{H}_{1}}{\gtrless}\gamma.\label{eq:LRT}
\end{equation}

We can conclude that as the number of pixels $N\rightarrow\infty$,
the CLT states that

\begin{eqnarray}
\Lambda^{\star}(\mathbf{e}) & = & \frac{\sum_{i=1}^{N}\Lambda(e_{i})-\mathbb{E}_{0}[\Lambda(e_{i})]}{\sqrt{\sum_{i=1}^{N}\mathrm{Var}_{0}[\Lambda(e_{i})]}}\label{eq:CLT}\\
 & \rightsquigarrow & \begin{cases}
\mathcal{N}(0,1) & \text{under }\mathcal{H}_{0}\\
\mathcal{N}(\delta,\varrho) & \text{under }\mathcal{H}_{1}
\end{cases},
\end{eqnarray}

where $\rightsquigarrow$ denotes the convergence in distribution,
$\delta$ is the deflection coefficient, and $\varrho$ is the effect
of embedding on the variance of the test statistic. We encourage the
reader to read the Appendix for more detailed information about the
test.

We can then compute the probability of detection $P_{\mathrm{D}}$
for a fixed False Alarm $P_{\mathrm{FA}}$:

\begin{eqnarray}
P_{\mathrm{D}} & = & Q\left(\frac{Q^{-1}(P_{\mathrm{FA}})-\delta}{\sqrt{\varrho}}\right).\label{eq:detection_probability}
\end{eqnarray}

In~\cite{Sed15MiPOD,Gib22TIFS} $\varrho=1$, which allows simply
to minimize the deflection coefficient $\delta$. In our setting,
unfortunately, the detection probability~\eqref{eq:detection_probability}
depends on the value $P_{\mathrm{FA}}$ since $\varrho\neq1$. Moreover,
$\delta$ and $\varrho$ are quite complicated functions of the change
rates, and it turns out to be numerically very unstable to try to
minimize $\delta/\sqrt{\varrho}$ or even $\delta$. For these reasons,
we will minimize an upper bound on the probability of detection instead.

\subsection{Kullback-Leibler Divergence}

\label{subsec:KLD}

From Sanov Theorem, we know that the probability of detection can
be upper bounded with

\begin{equation}
P_{\mathrm{D}}\leq1-e^{-\sum_{i=1}^{N}D(C_{i}||S_{i})},\label{eq:Sanov_bound}
\end{equation}
where $D(C_{i}||S_{i})$ is the Kullback-Leibler (KL) divergence between
$i$-th cover and stego pixels. Instead of minimizing the power of
the LRT, we will aim to minimize this upper bound, similarly to~\cite{Gib22si}.

We can now formalize our optimization problem as 
\begin{equation}
\min_{\boldsymbol{\beta}}\sum_{i=1}^{N}D(C_{i}||S_{i}),\label{eq:minimization_KLD}
\end{equation}
such that

\begin{equation}
\sum_{n=1}^{N}H_{3}(\beta_{n}^{+},\beta_{n}^{-})=\alpha,\label{eq:payload_constraint}
\end{equation}

where $\alpha$ is the desired relative payload and $H_{3}(\cdot,\cdot)$
is the ternary entropy function:

\begin{eqnarray}
H_{3}(\beta^{+},\beta^{-}) & = & -(1-\beta^{+}-\beta^{-})\log(1-\beta^{+}-\beta^{-})\nonumber \\
 &  & -\beta^{+}\log\beta^{+}-\beta^{-}\log\beta^{-}.\label{eq:ternary_entropy}
\end{eqnarray}
The objective~\eqref{eq:minimization_KLD} could be expressed as
minimizing the sum of KL divergences over every $8\times8$ block
of pixels, thanks to the block structure of JPEG images. We compute
the KL divergence for a single block of pixels as:

\begin{equation}
D(C||S)=\frac{1}{2}\sum_{i,j=0}^{7}\log\frac{\overline{\sigma}_{ij}^{2}}{\sigma_{ij}^{2}}+\frac{\mu_{ij}^{2}+\sigma_{ij}^{2}-\overline{\sigma}_{ij}^{2}}{\overline{\sigma}_{ij}^{2}}.\label{eq:KLD}
\end{equation}

Next, we simplify~\eqref{eq:KLD} by its second-degree Taylor polynomial
around $\boldsymbol{\beta}=\mathbf{0}$. The Taylor approximation
yields:

\begin{equation}
D(C||S)=\frac{1}{2}\sum_{k,l=0}^{7}\boldsymbol{\beta}_{kl}\mathbf{I}_{kl}\boldsymbol{\beta}_{kl}^{T},\label{eq:KLD_taylor}
\end{equation}

where $\boldsymbol{\beta}_{kl}=(\beta_{kl}^{+},\beta_{kl}^{-})$,
and 
\begin{equation}
\mathbf{I}_{kl}=\left[\begin{array}{cc}
I_{kl}^{+} & I_{kl}^{\pm}\\
I_{kl}^{\pm} & I_{kl}^{-}
\end{array}\right]\label{eq:FI_matrix}
\end{equation}
is the Fisher information matrix associated with $\boldsymbol{\beta}_{kl}$.
Let us denote 

\begin{eqnarray}
I_{kl} & = & q_{kl}^{4}\sum_{i,j=0}^{7}\frac{\left(f_{kl}^{ij}\right)^{4}}{\sigma_{ij}^{4}},\label{eq:Fisher_information}\\
\iota_{kl} & = & 2q_{kl}^{2}\sum_{i,j=0}^{7}\frac{\left(f_{kl}^{ij}\right)^{2}}{\sigma_{ij}^{2}}.\label{eq:Fisher_information_from_mean}
\end{eqnarray}

The entries of the Fisher information matrix~\eqref{eq:FI_matrix}
are given by:

\begin{eqnarray}
I_{kl}^{+} & = & (1-2u_{kl})^{4}I_{kl}+\iota_{kl},\label{eq:Fisher_informations_p}\\
I_{kl}^{-} & = & (1+2u_{kl})^{4}I_{kl}+\iota_{kl},\label{eq:Fisher_informations_m}\\
I_{kl}^{\pm} & = & (1+2u_{kl})^{2}(1-2u_{kl})^{2}I_{kl}-\iota_{kl}.\label{eq:Fisher_informations_pm}
\end{eqnarray}

We provide details on deriving the Fisher information matrix in the
Appendix.

Assuming $\sigma_{ij}^{2}\geq1$, the leading term of the Fisher information
is $\iota_{kl}$, which is present due to Eve's knowledge of $\boldsymbol{\mu}$.
Hence the embedding scheme derived by Alice is almost independent
of the side information. This makes intuitive sense because, for omniscient
Eve, it is much easier to estimate pixel mean (denoising) than to
estimate the variance. In turn, we could expect that Alice is limited
to very small embedding payloads.

\begin{figure}
\begin{centering}
\includegraphics[width=8cm]{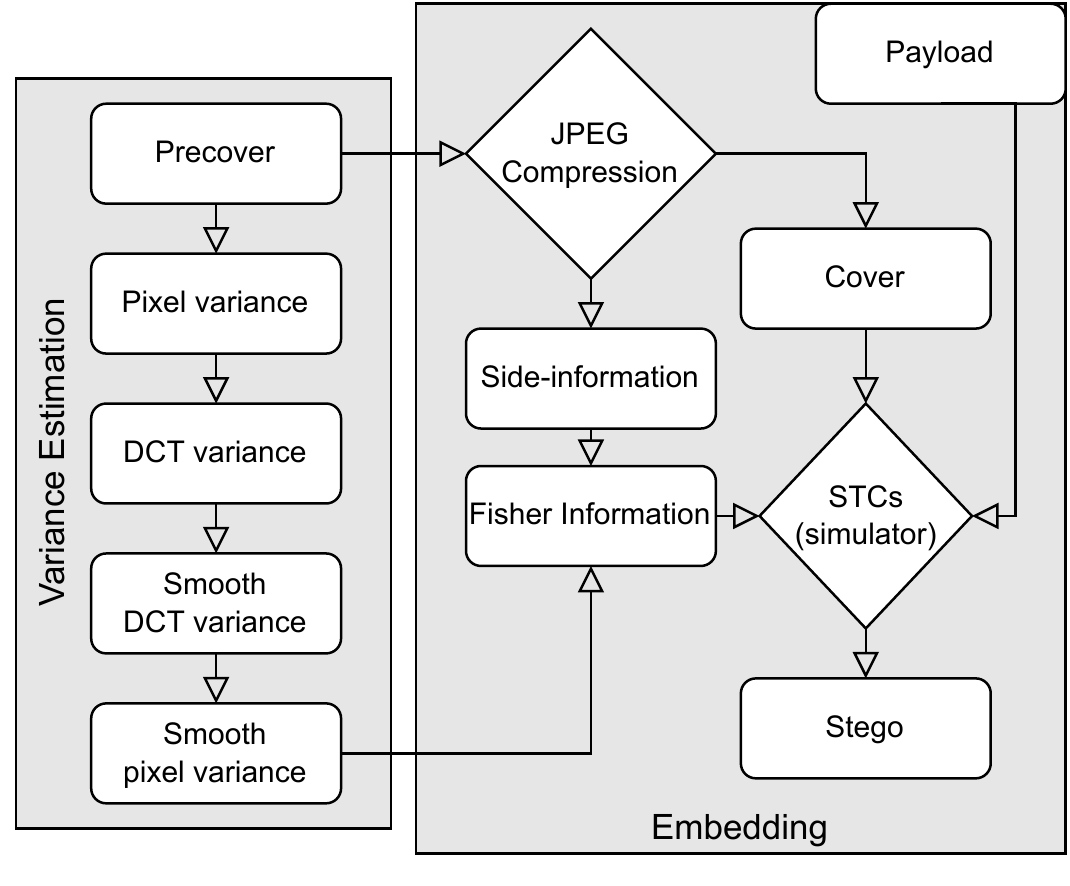}
\par\end{centering}
\caption{\label{fig:embedding_pipeline}Embedding pipeline of JEEP. The Variance
Estimation branch could be optionally replaced by a different methodology.}
\end{figure}

\subsection{Realistic Attacker}

\label{subsec:Realistic-Attacker}

As mentioned in the previous section, full knowledge of the side information
virtually disables its effect during embedding. In this section, we
thus consider more realistic (even though still very powerful) Eve.
In particular, we assume Eve has access to the variance of embedding
changes $\mathrm{Var}(\eta)$ and the covariance between the embedding
changes and the rounding errors $\mathrm{Cov}(\eta,U)$. On the other
hand, we assume that Eve does not have access to the precover or the
expectation of embedding change $\mathbb{E}[\eta]$. In practice,
we are not sure if Eve can access the residuals $\mathbf{e}$, but
it was shown~\cite{But20rjca} that the decompression rounding errors
$\mathbf{y}-[\mathbf{y}]$ are excellent approximations of such signals.\footnote{At least for quality factors 99 and 100.}
For this reason, we assume that Eve does, in fact, have access to
the residuals. It is helpful to mention that with the knowledge we
granted Eve, she cannot reconstruct the change rates $\beta_{kl}^{\pm}$
nor the side information $u_{kl}$. That is in accordance with our
real-world expectations since the change rates are correlated with
the side information, which is, by definition, inaccessible.

Eve's hypothesis test then transforms into

\begin{eqnarray}
\mathcal{H}_{0} & : & \mathbf{e}\sim\mathcal{N}(0,\Sigma_{ij}),\label{eq:null_hypothesis_pixels_constrained}\\
\mathcal{H}_{1} & : & \mathbf{e}\sim\mathcal{N}(0,\overline{\Sigma}_{ij}).\label{eq:alt_hypothesis_pixels_constrained}
\end{eqnarray}

Following the same methodology from the previous section, we find
that Alice wants to minimize the KL divergence~\eqref{eq:KLD_taylor},
but the elements of the Fisher information matrix~\eqref{eq:FI_matrix}
are of the form

\begin{eqnarray}
I_{kl}^{+} & = & (1-2u_{kl})^{4}I_{kl},\label{eq:FI_p}\\
I_{kl}^{-} & = & (1+2u_{kl})^{4}I_{kl},\label{eq:FI_m}\\
I_{kl}^{\pm} & = & (1+2u_{kl})^{2}(1-2u_{kl})^{2}I_{kl},\label{eq:FI_pm}
\end{eqnarray}

with $I_{kl}$ from~\eqref{eq:Fisher_information}.

The proposed strategy assumes that pixel variance is known to the
steganographer. However, this is not the case in practice because
Alice has an image and needs to estimate the variance in some way
before the embedding procedure (unless we assume Alice has additional
knowledge, which we do not in this work). Similarly, Eve, who observes
a potential stego version of the same image, cannot know the actual
variance of pixels. Since we want to compare two embedding strategies,
assuming omniscient or realistic attacker, and we want to avoid the
effect of imprecise variance estimation, we create an artificial source
where we have control over pixel variances. This source is detailed
in Section~\ref{subsec:Controlled-Source-dataset}, and Section~\ref{subsec:Attacker_effect}
shows that assuming omniscient Eve compared to the realistic one leads
to severe security decay when tested against a state-of-the-art steganalyzer.

\subsection{Variance Estimation}

\label{subsec:Variance-estimation}

In an image source where we do not know the variance of the noise,
it is necessary to estimate it. However, this could create errors
due to imprecise estimation. We will therefore use some of the best
practices for estimating the variance. We consider two different variance
estimators for two different situations. The first is for steganography
targeting 'standard' steganalysis, while the second targets the Reverse
JPEG Compatibility Attack used for images compressed with QF 100.

\subsubsection{MiPOD-Estimator}

For 'standard' steganography, the methodology of computing variance,
depicted in Figure~\ref{fig:embedding_pipeline}, is described in
the following. First, we need to estimate the pixel variances $\sigma_{ij}^{2}$.
We decided to use a popular trigonometric variance estimator originally
used for MiPOD~\cite{Sed15MiPOD}, which is also used in its JPEG
extension~\cite{Cog20color,Gib22TIFS}. For the sake of brevity,
we refer the reader to the publications mentioned above for more details
about the variance estimator.

Due to imprecise variance estimates, smoothing with the neighboring
blocks in the DCT domain, as proposed in~\cite{Gib22TIFS}, is done
to mitigate overly-content adaptive embedding. First, we compute the
DCT variances $\Sigma_{kl}=\mathbf{D}\Sigma_{ij}\mathbf{D}^{T}$.

The squared variances are then smoothed by averaging nine neighboring
DCT coefficients from the same DCT mode:

\begin{equation}
\tilde{\sigma}_{kl}^{-4}=\sum_{i=-1}^{1}\sum_{j=-1}^{1}w_{ij}\sigma_{k+8i,l+8j}^{-4},\label{eq:DCT_variance_smoothing}
\end{equation}

where $\mathbf{w}$ is the averaging kernel:

\begin{equation}
\mathbf{w}=\frac{1}{20}\left(\begin{array}{ccc}
1 & 3 & 1\\
3 & 4 & 3\\
1 & 3 & 1
\end{array}\right).\label{eq:averaging_kernel}
\end{equation}

Finally, the smooth DCT variances are lower-bounded by $\tilde{\sigma}_{kl}^{2}=\min\{10^{-10},\tilde{\sigma}_{kl}^{2}\}$,
to prevent numerical instabilities, and decompressed to obtain final
smoothed variance estimates $\tilde{\Sigma}_{ij}=\mathbf{D}^{T}\tilde{\Sigma}_{kl}\mathbf{D}$.

While we think there are many possible ways of smoothing the variance,
we chose this (perhaps cumbersome) way because it has been shown in
practice that smoothing the DCT variances provides good results for
steganography.

\begin{figure}
\begin{centering}
\includegraphics[width=4cm]{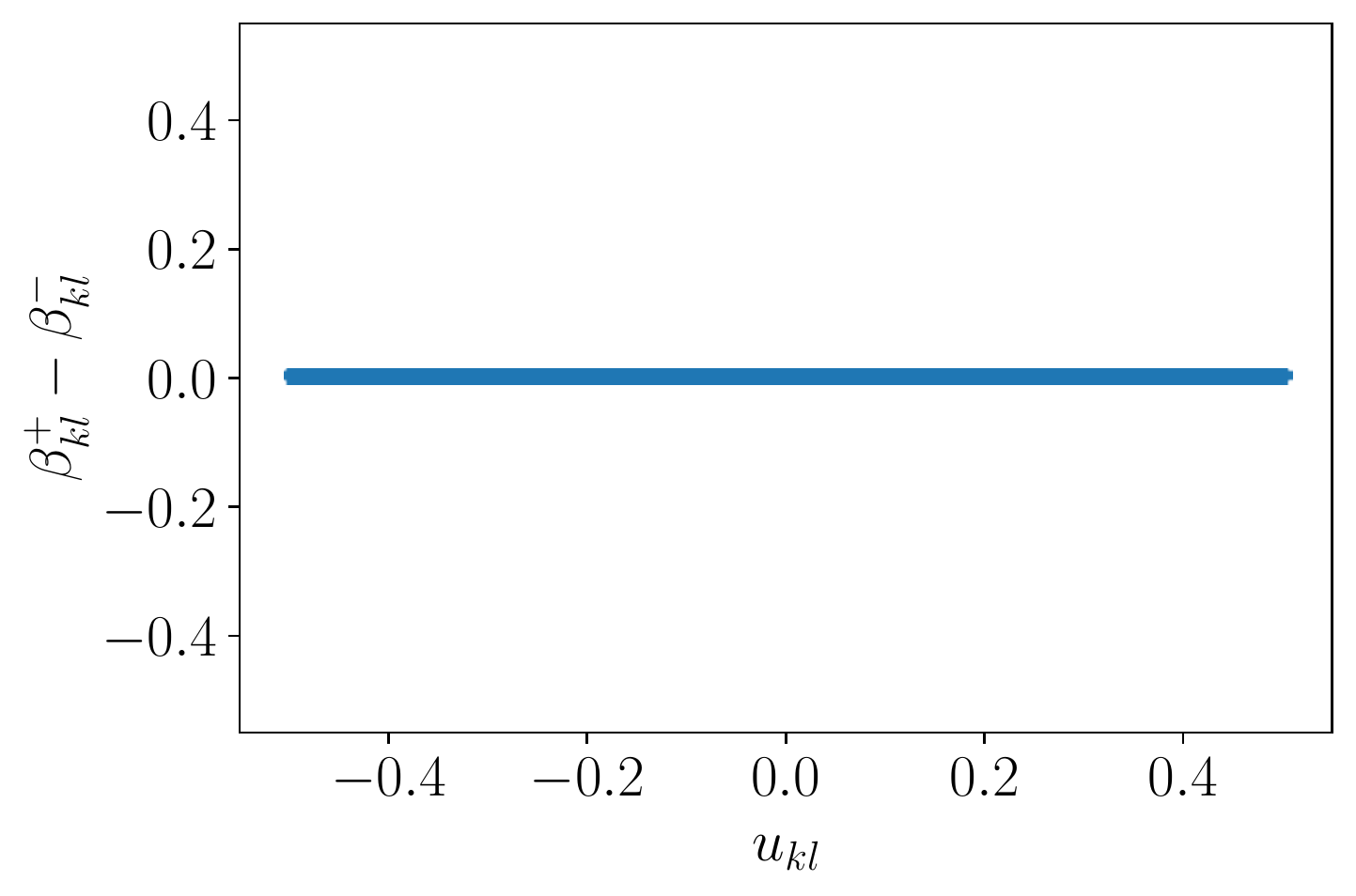}\includegraphics[width=4cm]{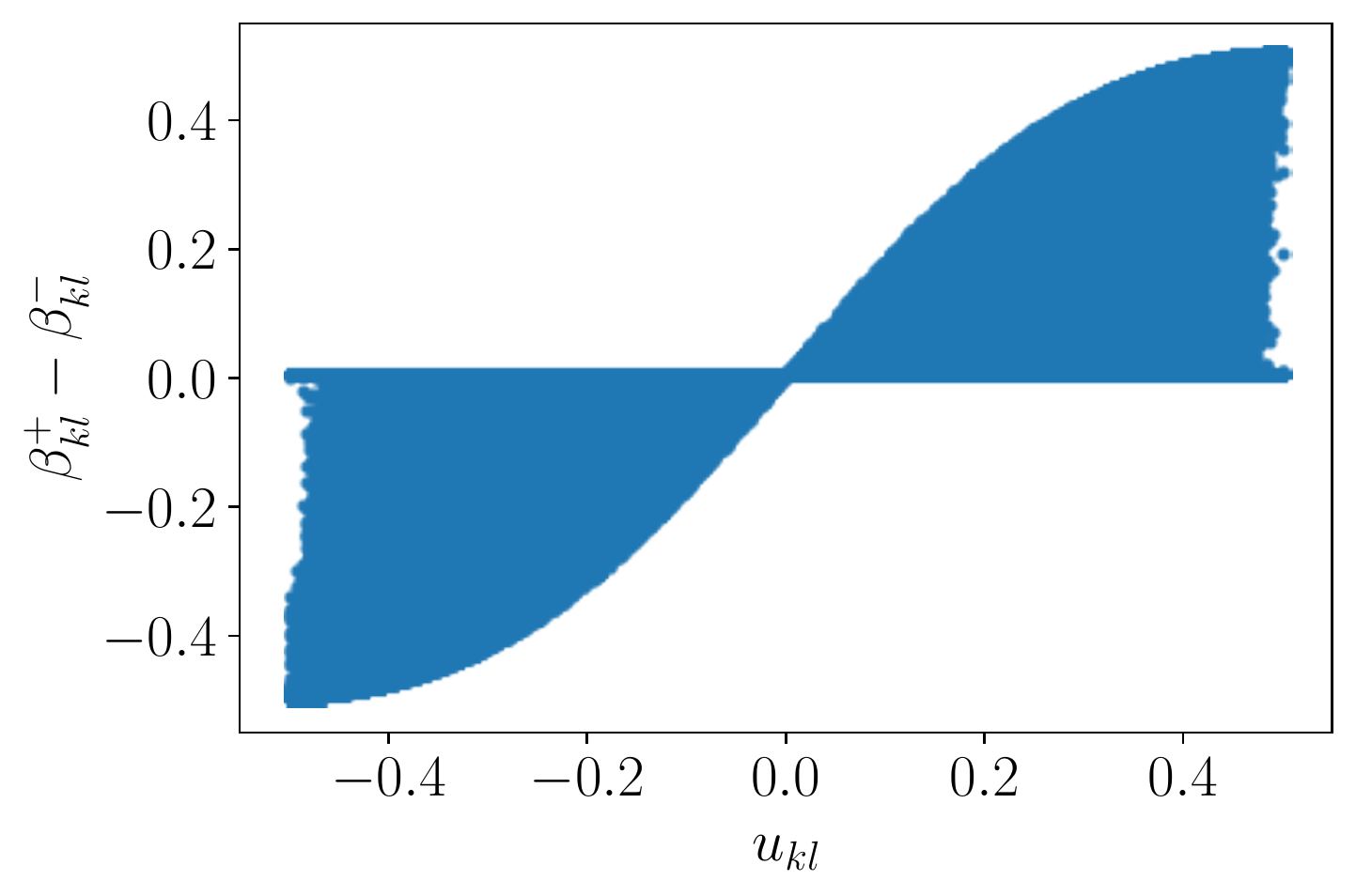}
\par\end{centering}
\caption{\label{fig:changes_vs_sideinformation}Expectation of embedding change
as a function of the rounding error $u_{kl}$ from one image, QF 95,
0.5 bpnzAC. Left: JEEP(o), right: JEEP(r).}
\end{figure}

\subsubsection{RJCA-Estimator}

The second variance estimator we use in this work is for images compressed
with QF 100. A different variance estimator is necessary because RJCA~\cite{But20rjca}
can be used to steganalyze these images, and the MiPOD estimator was
designed with feedback from spatial domain steganalysis. We decided
to use constant variance for all pixels since it was shown~\cite{But22SVP}
that using constant costs with a cleverly picked polarity of the embedding
changes is a good strategy against the RJCA. We refer to JEEP with
this constant variance estimator as JEEP-C. Unlike in the previous
section, JEEP-C does not require Fisher Information smoothing. We
also tested the MiPOD variance estimator for this case (JEEP); however,
its performance was worse than using simple constant variance, and
thus we do not mention its results.

\subsection{Obtaining the Change Rates}

\label{subsec:Obtaining-change_rates}

The goal of the proposed scheme is to solve~\eqref{eq:minimization_KLD}
under the payload constraint~\eqref{eq:payload_constraint}. To achieve
this, we use the method of Lagrange multipliers. Together with the
payload constraint~\eqref{eq:payload_constraint}, this forms $2N+1$
equations with $2N+1$ unknowns composed of the change rates $\beta_{n}^{\pm},\,n=1,\ldots,N$,
and the Lagrange multiplier $\vartheta$:

\begin{eqnarray}
\beta_{n}^{+}I_{n}^{+}+\beta_{n}^{-}I_{n}^{\pm} & = & \lambda\log\frac{1-\beta_{n}^{+}-\beta_{n}^{-}}{\beta_{n}^{+}},\label{eq:solving_plus}\\
\beta_{n}^{-}I_{n}^{-}+\beta_{n}^{+}I_{n}^{\pm} & = & \lambda\log\frac{1-\beta_{n}^{+}-\beta_{n}^{-}}{\beta_{n}^{-}},\label{eq:solving_minus}\\
\sum_{n=1}^{N}H_{3}(\beta_{n}^{+},\beta_{n}^{-}) & = & \alpha.\label{eq:solving_payload}
\end{eqnarray}

Equations~\eqref{eq:solving_plus}-\eqref{eq:solving_payload} can
be solved efficiently with the Newton method by parallelizing over
all $N$ DCT coefficients and a binary search over $\lambda$ satisfying
the payload constraint.

To employ a coding mechanism, such as Syndrome-trellis codes~\cite{Fil10stc},
the change rates can be converted into embedding costs:

\begin{eqnarray}
\rho_{n}^{\pm} & = & \log\left(\frac{1-\beta_{n}^{+}-\beta_{n}^{-}}{\beta_{n}^{\pm}}\right).\label{eq:costs}
\end{eqnarray}

The costs are obtained by inverting the formula for optimal change
rates, given embedding costs:

\begin{equation}
\beta_{n}^{\pm}=\frac{e^{-\lambda\rho_{n}^{\pm}}}{1+e^{-\lambda\rho_{n}^{+}}+e^{-\lambda\rho_{n}^{-}}}.\label{eq:optimal_change_rates}
\end{equation}

\subsection{Discussion}

\label{subsec:Discussion}

We observed that constraining embedding in the 'rational' DCT modes,
as described in Section~\ref{subsec:SI_literature}, is indeed necessary
in order to avoid security deterioration. However, we can see that
this would not be the case for omniscient Eve since then the term~\eqref{eq:Fisher_information_from_mean}
needs to be present in the Fisher information, which automatically
avoids the problem of Fisher information being zero.

It was shown~\cite{But20ih} that steganographic costs can be viewed
as estimators of a reciprocal standard deviation of pixels. Seeing
the linear relationship between the standard deviation and the side
information~\eqref{eq:FI_p},\eqref{eq:FI_m} could explain why using
the linear term $(1-2u_{kl})$ in the cost-based side-informed steganography~\eqref{eq:rho_plus_si2},\eqref{eq:rho_minus_si2}
works best in practice. Interestingly, a second power of $(1-2u_{kl})$~\eqref{eq:FI_si}
was derived in~\cite{Den17si} from the image model. We believe this
is due to modeling stego image as a mixture of Gaussian distributions.
The fourth power of $(1\pm2u_{kl})$ present in~\eqref{eq:FI_p},\eqref{eq:FI_m}
was already used in SI-JMiPOD~\cite{Gib22TIFS}, but the explanation
was heuristic and did not come from the image model. This heuristic
also prevented the authors from using ternary embedding since they
were probably unsure how to handle the term $I_{kl}^{\pm}$. Since
JMiPOD models stego DCT coefficients as a Gaussian mixture, we believe
the modulation~\eqref{eq:FI_si} should have been used instead, similarly
as in SI-MiPOD~\cite{Den17si}. 

Lastly, if the side information is unavailable, JEEP naturally degenerates
into a non-informed scheme, simply by setting $\mathbb{E}[u]=0$.
Interestingly, even though the proposed method changes pixel variance
instead of mean, the resulting Fisher information closely resembles
that of JMiPOD. While JMiPOD computes DCT variance~\eqref{eq:spatial_to_dct_var}
and uses its square to compute the Fisher information $\sigma_{kl}^{-4}=\left(\sum_{i,j=0}^{7}\left(f_{kl}^{ij}\right)^{2}\sigma_{ij}^{2}\right)^{-2}$,
we found out that the Fisher information contains $\sum_{i,j=0}^{7}\frac{\left(f_{kl}^{ij}\right)^{4}}{\sigma_{ij}^{4}}$
instead. It can be easily shown that the value computed by JMiPOD
is always greater than $I_{kl}$, making the embedding more content-adaptive.

\begin{figure}
\begin{centering}
\includegraphics[width=8cm]{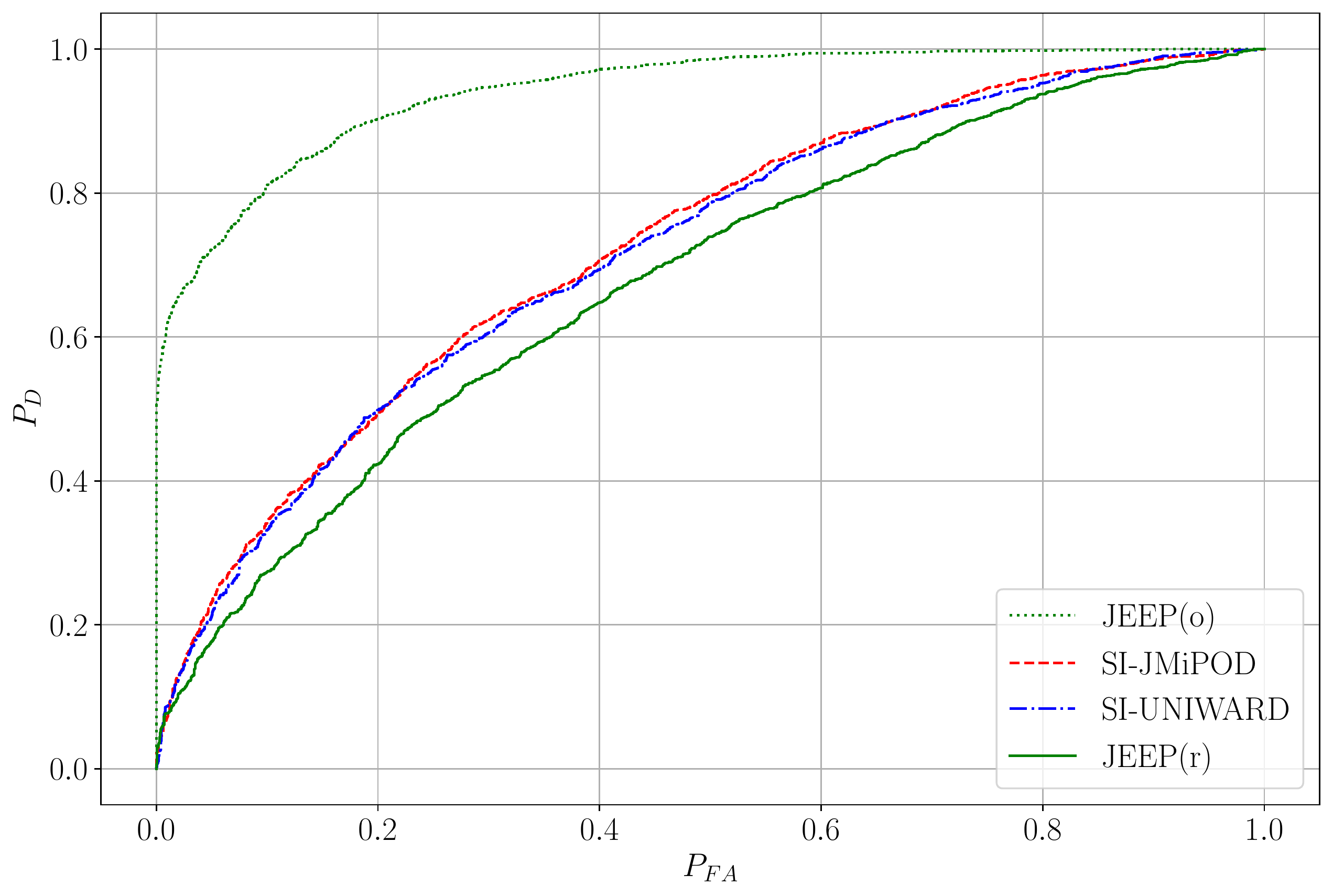}
\par\end{centering}
\caption{\label{fig:ROC_artificial}ROC curves of JIN-SRNet, trained on N-BOSSBase
cover images compressed with QF 95 and stego images embedded with
0.5 bpnzAC.}
\end{figure}

\begin{figure*}
\begin{centering}
\includegraphics[viewport=50bp 10bp 780bp 390bp,clip,width=8cm]{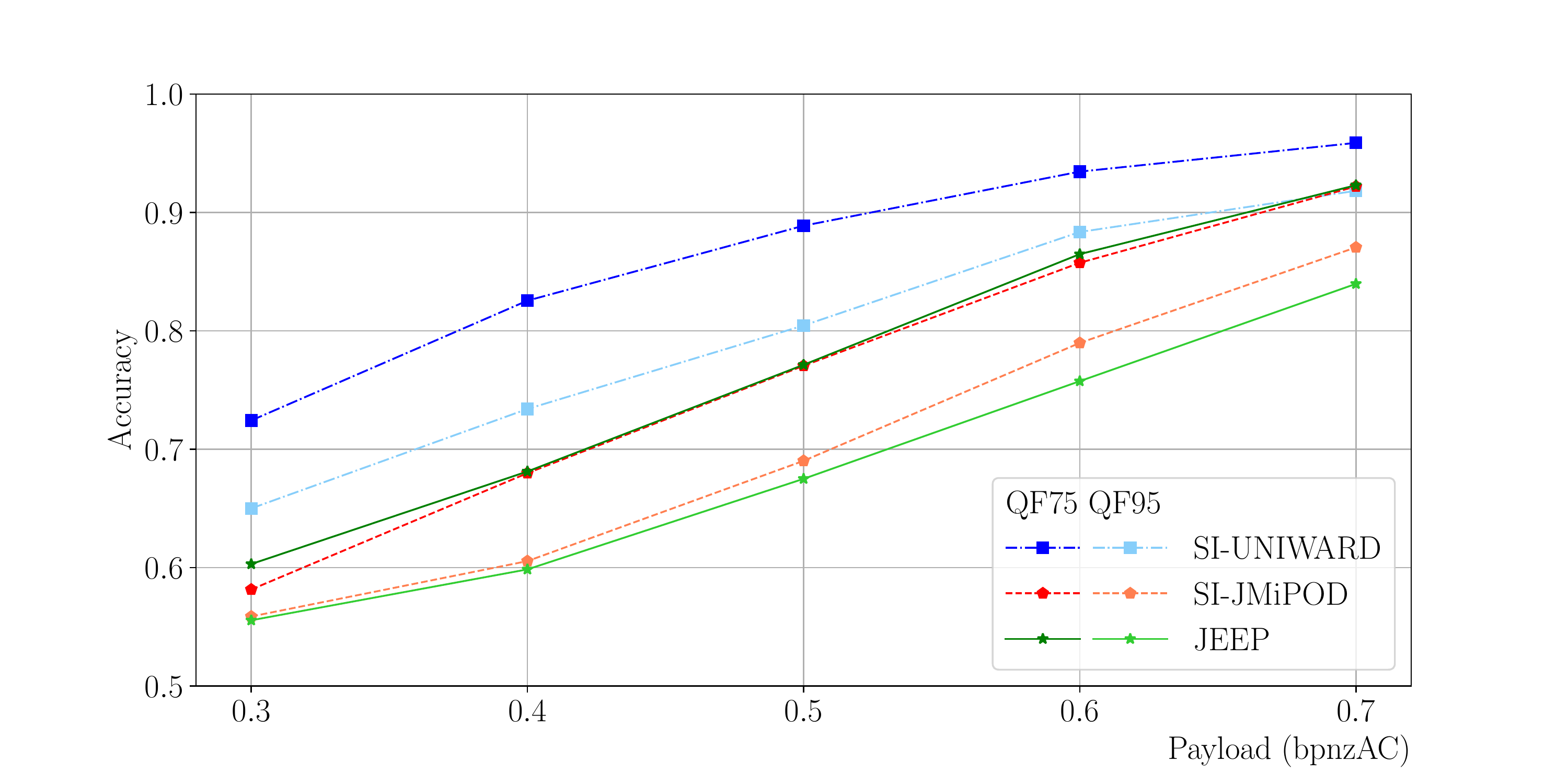}\hspace{1cm}\includegraphics[viewport=50bp 10bp 780bp 390bp,clip,width=8cm]{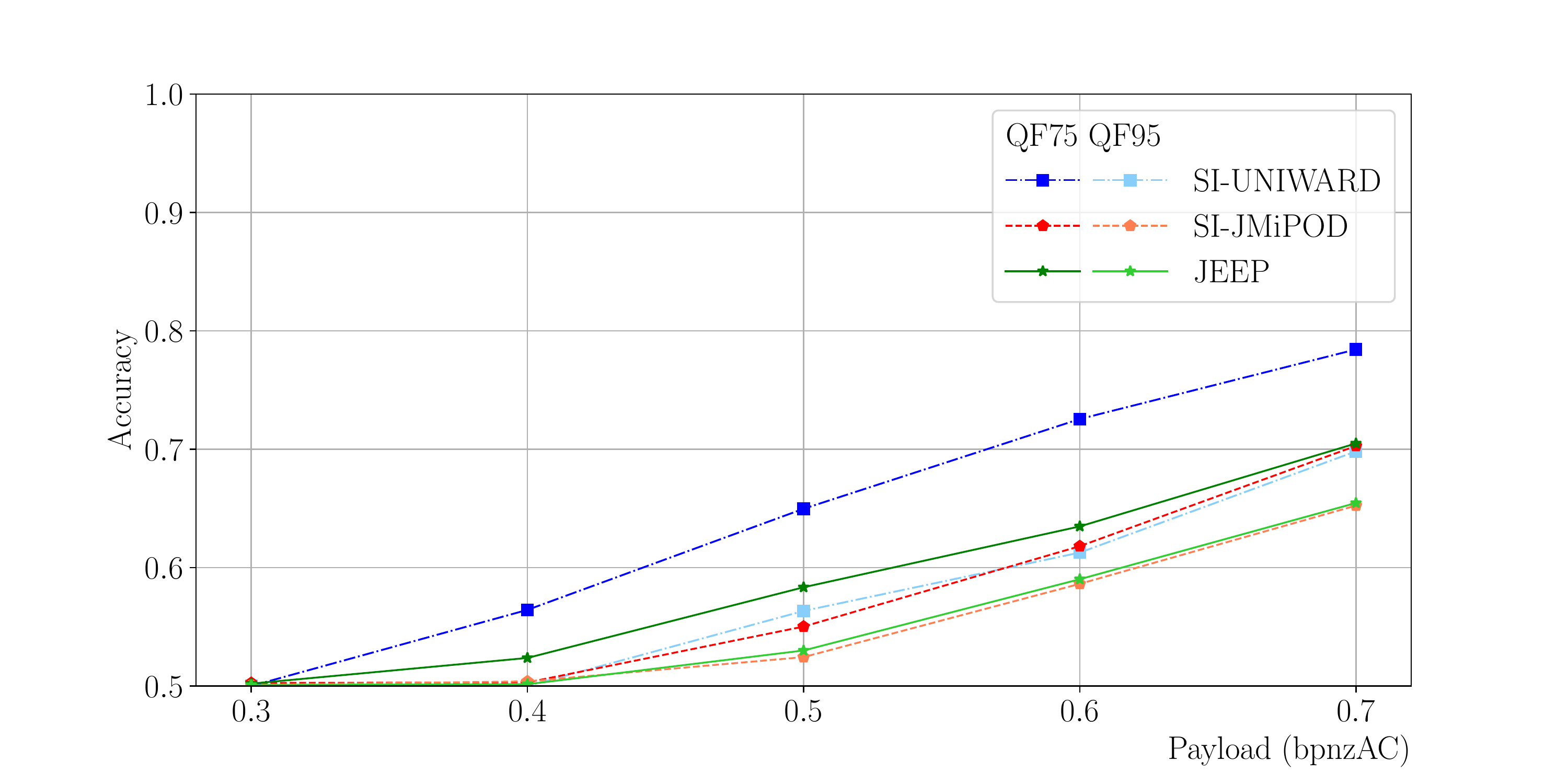}
\par\end{centering}
\begin{centering}
\par\end{centering}
\caption{\label{fig:Accuracy_BB_spatial}Accuracy of JIN-SRNet (left) and DCTR
(right) for BOSSBase.}
\end{figure*}

\begin{figure}
\begin{centering}
\includegraphics[width=8cm]{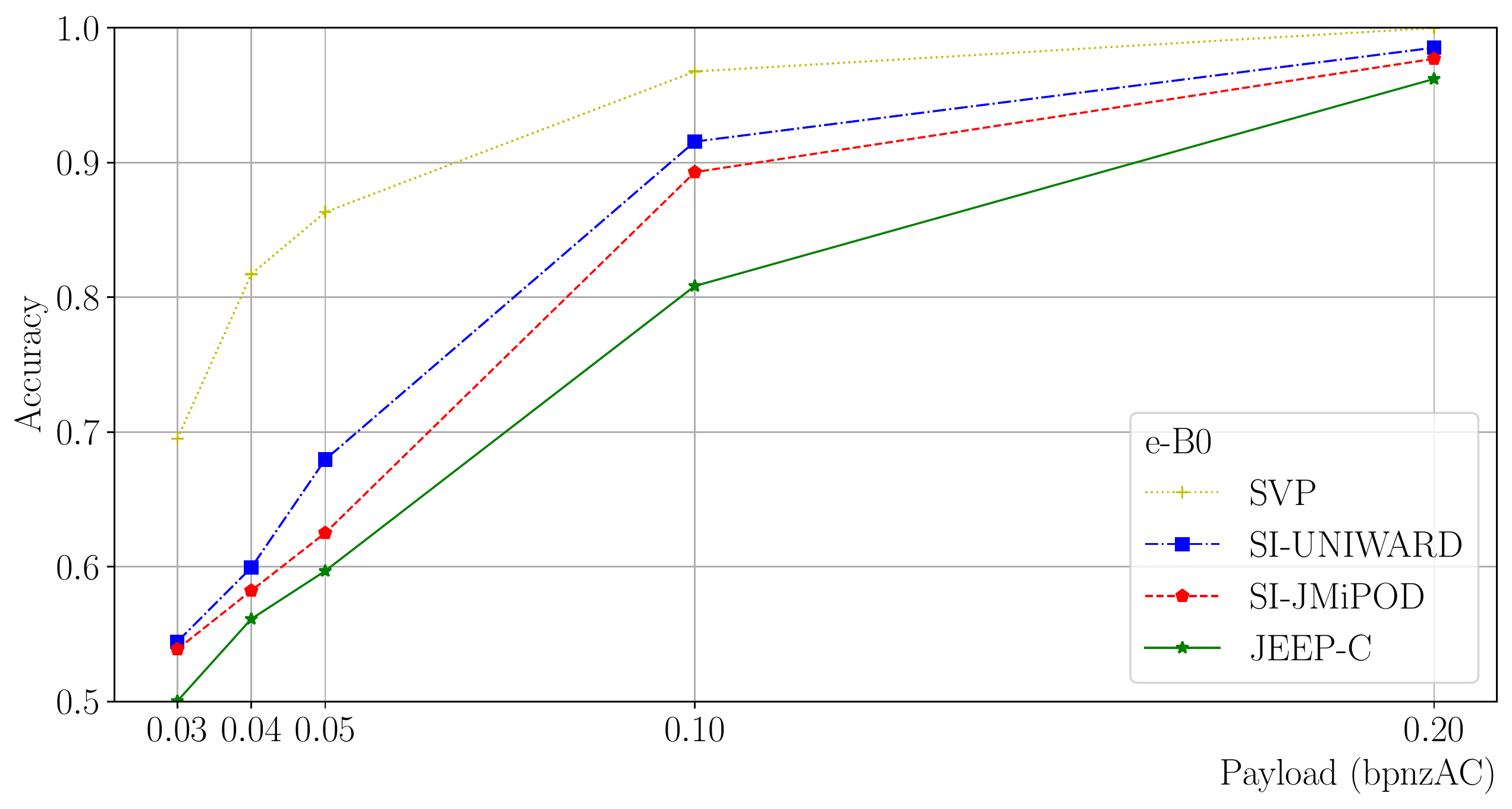}
\par\end{centering}
\caption{\label{fig:Accuracy_BB_RJCA}Accuracy of e-B0 for BOSSBase images
compressed with QF 100.}
\end{figure}

\section{Datasets and Detectors}

\label{sec:Dataset-and-detectors}

To experimentally verify our results, we chose two popular datasets
used for steganographic benchmarking. First is the BOSSbase~\cite{Bas11BOSS}
dataset, made of $10,000$ uncompressed grayscale images of size $512\times512$,
split into training, validation, and testing sets of sizes $7,000$,
$1,000$, and $2,000$, respectively. The other dataset is the ALASKA2
dataset~\cite{cogranne20wifs}, comprising $25,000$ uncompressed
grayscale images of size $512\times512$. We split the ALASKA2 dataset
into training, validation, and testing sets of $22,000$, $1,000$,
and $2,000$ images. Both datasets are then JPEG compressed with python's
PIL library with several quality factors (QF). For security comparison
with the prior art, we picked SI-UNIWARD~\cite{Den15wifs,Hol13eur,Hol13ih}
and SI-JMiPOD~\cite{Gib22TIFS} algorithms as two main representatives
of state-of-the-art steganography with side information. To add more
contrast w.r.t. prior art on QF 100, we added SVP~\cite{But22SVP}.
This algorithm was explicitly designed to be robust against the RJCA,
even though without the use of side information.

We use three types of detectors to evaluate steganographic security.
First is EfficientNet-B0~\cite{Mingxing19efficientnet}, initialized
with weights pre-trained on the ImageNet dataset~\cite{deng09imagenet}.
This detector, which we denote as e-B0, is used to test security against
the RJCA and is trained only on the spatial domain rounding errors
$e_{ij}=y_{ij}-[y_{ij}]$ of images compressed with QF 100. The second
detector we use to verify security in the pixel domain is the JIN-SRNet~\cite{But21JIN},
SRNet~\cite{Bor18} pre-trained on ImageNet embedded with J-UNIWARD
at a variable payload. The SRNet was trained with the Pair Constraint
(PC) - forcing the cover and its stego version into the same mini-batch.
Unlike in~\cite{But21JIN}, it was observed in this work to boost
the network's performance. We believe this is caused by a much bigger
security of the side-informed algorithms. The PC was not used for
the e-B0, as it did not bring any detection improvements. Mini-batch
size of both detectors was set to 32 images. The first detector is
only trained for 15 epochs, as the detection saturates quickly in
this scenario. The JIN-SRNet, on the other hand, is trained for 50
epochs in BOSSBase and 20 epochs in ALASKA2 due it its much bigger
training set. Because the detectability in the spatial domain is typically
much smaller than in the RJCA scenarios, we use larger payloads for
the spatial domain detector. The last detector we use is the Low-Complexity
Linear Classifier~\cite{Cog15wifs} (LCLC) coupled with DCTR features~\cite{Hol14dctr}.
We include this detector to verify security in the DCT domain as well.
The LCLC was trained on half of the images and tested on the other
half for both datasets.

\subsection{Controlled Source}

\label{subsec:Controlled-Source-dataset}

To verify the validity of our assumptions on Eve's knowledge, we created
an artificial cover source~\cite{Bor19ei}, which allows us to have
the true pixel variances without the need to estimate them. In this
dataset, we denoise the images to eliminate the dependencies caused
by the RAW development and further processing. The images are then
noisified to enforce our cover model~\eqref{eq:cover_distribution}.
We briefly summarize the creation of the dataset in 5 steps: 

1) Estimate pixel variance $\sigma_{ij}^{2}$ using MiPOD's variance
estimator introduced in Section~\eqref{subsec:Variance-estimation}. 

2) Denoise every image with Daubechies 8-tap wavelets~\cite{Mih99a}
by removing i.i.d. Gaussian noise with a standard deviation $\sigma_{\mathrm{den}}=10$.
The (non-integer) pixel values of the denoised image are clipped to
the dynamic range of 8-bit grayscale images $[0,255]$. 

3) Narrow the dynamic range into $[15,240]$ by a linear transformation
of the pixels and round them to integers. Denote them as $\mu_{ij}$.

4) Adjust the variance so that the probability of a noisified pixel
being outside of the interval $[0,255]$ is equal to a one-sided $5\sigma$
Gaussian outlier ($2.87\times10^{-7}$). This is done by computing
$\underline{\sigma}_{ij}=\min\{\frac{1}{5}\min\{\mu_{ij},255-\mu_{ij}\},\sigma_{ij}\}$. 

5) Noisify cover pixels $x_{ij}$ by adding samples from $\mathcal{N}(0,\underline{\sigma}_{ij}^{2})$.
The resulting cover image model comprises independent Gaussian variables
$\mathcal{N}(x_{ij},\underline{\sigma}_{ij}^{2})$ rounded to the
nearest integers and clipped to $[0,255]$. 

For more details about the 'noisifying' procedure, see~\cite{Bor19ei}.
This noisified dataset was created from BOSSbase, and we refer to
it as N-BOSSbase.

\begin{figure*}
\begin{centering}
\includegraphics[viewport=50bp 10bp 780bp 390bp,clip,width=8cm]{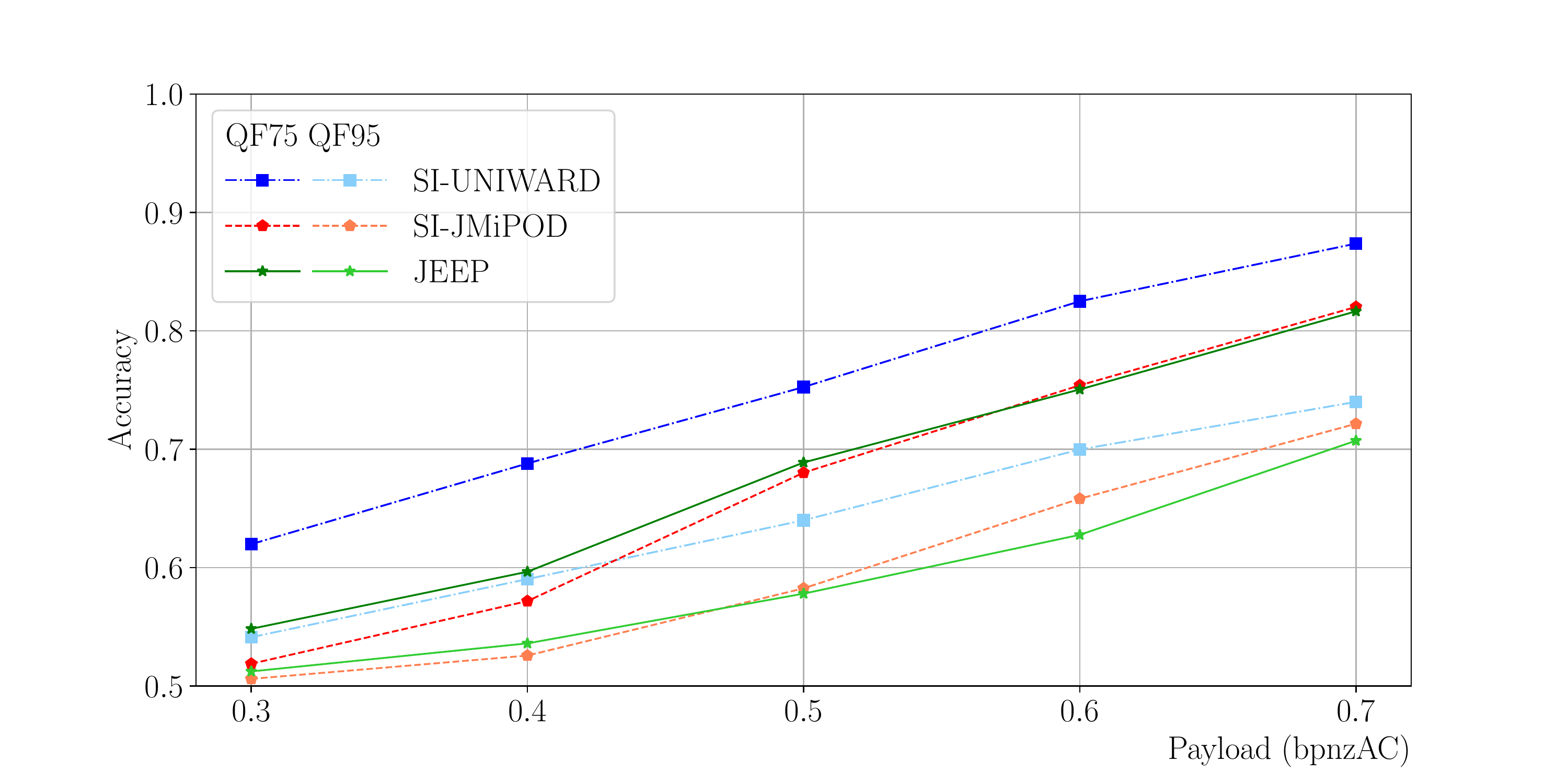}\hspace{1cm}\includegraphics[viewport=50bp 10bp 780bp 390bp,clip,width=8cm]{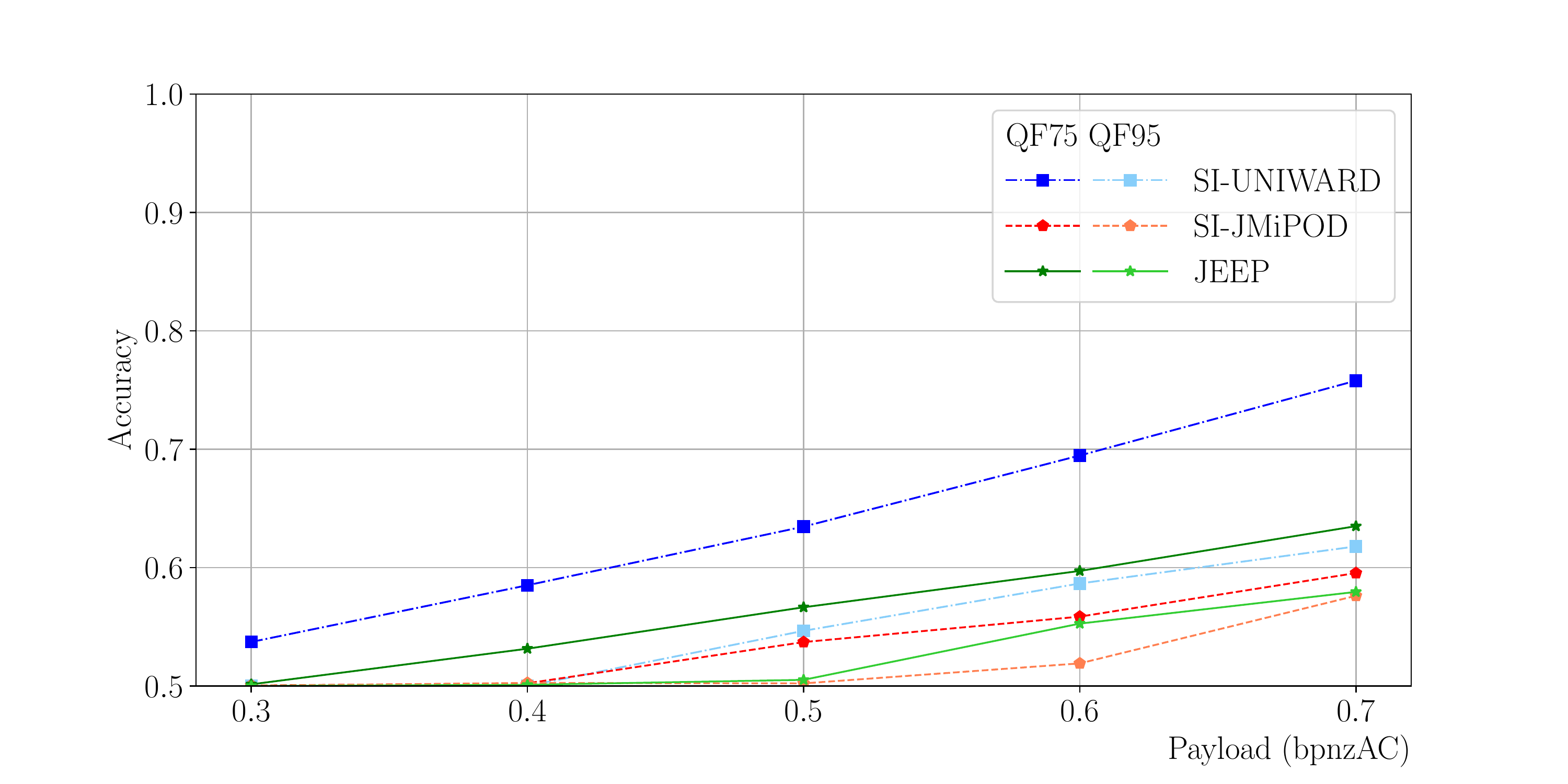}
\par\end{centering}
\caption{Accuracy of JIN-SRNet (left) and DCTR (right) for ALASKA2.}
\end{figure*}

\begin{figure}
\begin{centering}
\includegraphics[width=8cm]{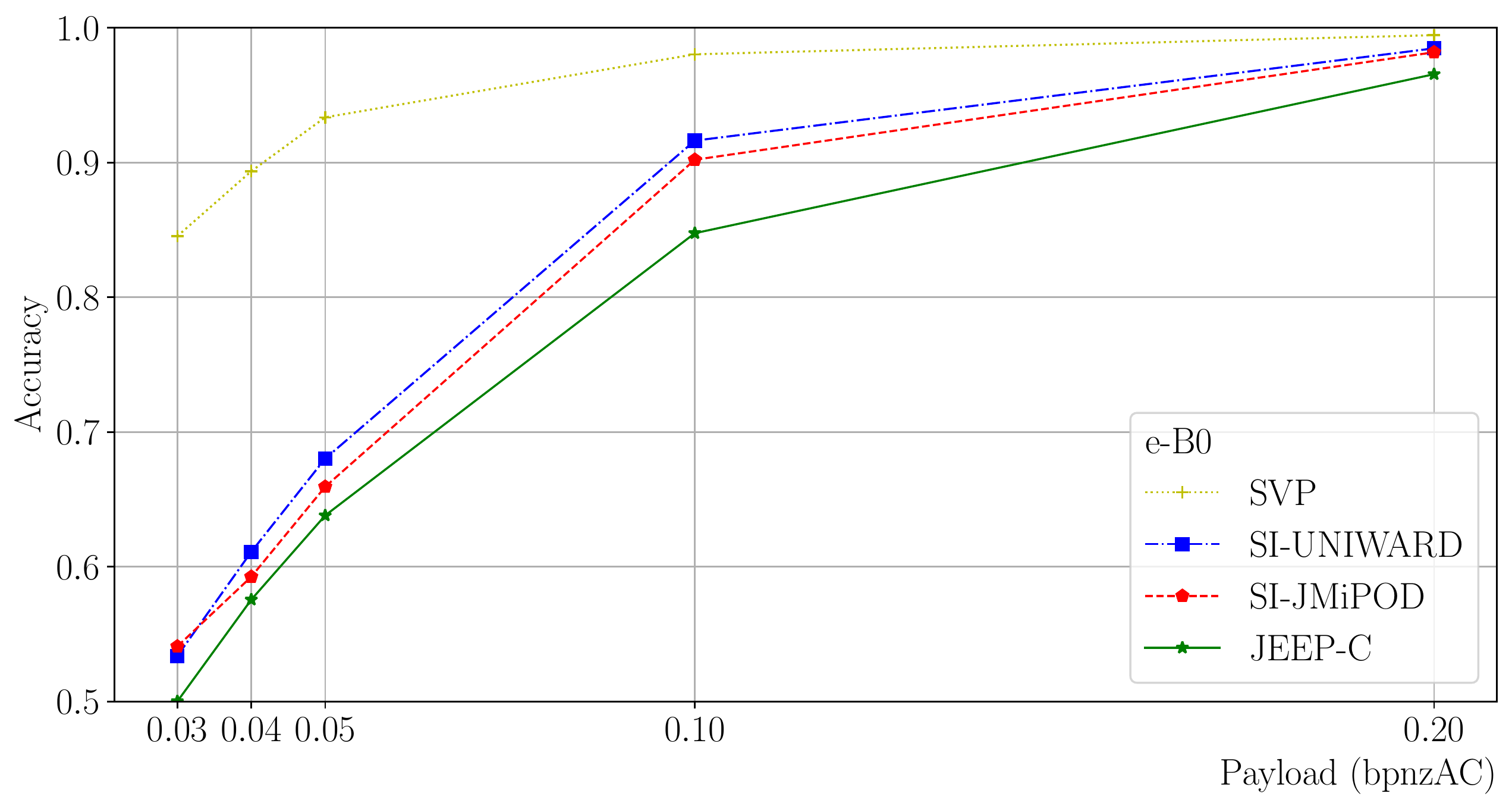}
\par\end{centering}
\caption{\label{fig:Accuracy_A2_RJCA}Accuracy of eB0 for ALASKA2 images compressed
with QF 100.}
\end{figure}

\section{Results}

\label{sec:Results}

In this section, we first evaluate the validity of assumptions on
the steganalyst by training a detector in a controlled environment.
We show that assuming all-powerful Eve is a fundamental mistake Alice
can make because the detectability of her resulting embedding algorithm
is extremely high. In the other two sections, we evaluate JEEP in
BOSSBase and ALASKA2 datasets with several steganalyzers. We show
that it provides superior security to previous state-of-the-art side-informed
steganography.

\subsection{Attacker Capabilities Effect (Controlled Source)}

\label{subsec:Attacker_effect}

In this section, we aim to investigate the difference between Alice's
assumptions on Eve. On the one hand, Eve is assumed to be omniscient
- has full access to the side information. We have shown in Section~\ref{subsec:KLD}
that Alice's embedding strategy, in this case, is to minimize the
KL divergence with the Fisher information matrix~\eqref{eq:Fisher_informations_p}-\eqref{eq:Fisher_informations_pm}.
On the other hand, Alice can assume a more realistic Eve described
in Section~\ref{subsec:Realistic-Attacker}. The only thing Alice
needs to change in such a scenario is using the Fisher information
matrix of the form~\eqref{eq:FI_p}-\eqref{eq:FI_pm} instead. We
refer to these two embedding algorithms as JEEP(o) and JEEP(r) for
omniscient and realistic. 

In Figure~\ref{fig:changes_vs_sideinformation}, we show the expectation
of embedding change $\beta^{+}-\beta^{-}$ as a function of the rounding
error $u$ for the omniscient and realistic attacker. We can clearly
see that for the omniscient Eve, the side information has no effect
on the embedding because it stays symmetric. For the realistic attacker,
the side information introduces a lot of asymmetry.

Next, we trained JIN-SRNet on the denoised images from N-BOSSBase
compressed with QF 95 and embedded with $0.5$ bits per non-zero AC
DCT coefficient (bpnzAC). We used SI-JMiPOD, SI-UNIWARD, and JEEP(o,r).
Since the pixel variances are known for all methods, we skipped the
smoothing~\eqref{eq:DCT_variance_smoothing} (SI-UNIWARD does not
use any smoothing). We show the ROC curves of these detectors in Figure~\ref{fig:ROC_artificial}.
Several observations can be made from this figure. First, the assumption
of omniscient Eve during embedding makes, in fact, makes the scheme
highly detectable. Next, JEEP(r) provides the best security across
embedding schemes. We believe the improvement over SI-JMiPOD comes
mainly from two things. 1) JEEP is, by nature, a ternary embedding
scheme, which reduces the total number of embedding changes. 2) Different
methodology of computing the Fisher information aimed specifically
against spatial steganalysis, as pointed out in Section~\ref{subsec:Discussion}.
And lastly, SI-JMiPOD and SI-UNIWARD have a very similar ROC curve,
which is somewhat surprising because in uncontrolled datasets (see
Sections~\ref{subsec:results_boss},\ref{subsec:results_ALASKA}),
SI-UNIWARD is way more detectable.

We have seen that with a state-of-the-art steganalyzer, assuming an
omniscient attacker during embedding leads to severe security underperformance.
That makes us conclude that such an assumption is unrealistic, and
for the rest of the paper, we only assume realistic Eve. We will refer
to Alice's embedding algorithm simply as JEEP.

\subsection{BOSSBase}

\label{subsec:results_boss}

In this section, we evaluate the empirical security of JEEP in the
popular BOSSBase database. In Figure~\ref{fig:Accuracy_BB_spatial},
we plot the accuracies of SRNet and DCTR as a function of relative
payload $\alpha$ for QFs 75 and 95. When steganalyzing with SRNet,
we can see similar behavior for JEEP and SI-JMiPOD at QF 75, with
SI-UNIWARD being much more detectable across all payloads. For QF
95, we can see better security for JEEP by up to 3.2\% in terms of
accuracy. With DCTR, we can see that JEEP is more detectable than
SI-JMiPOD. That is not surprising, as JEEP was built with a decompressed
image model, while JMiPOD uses a model in the DCT domain, where the
steganalysis is performed. However, we can see that the detectability
is much higher in the spatial domain across all QFs and payloads.

Results for images compressed with QF 100 are shown in Figure~\ref{fig:Accuracy_BB_RJCA}.
Notice that the payloads are much smaller due to the detection power
of RJCA. While JEEP is slightly more secure than SI-JMiPOD (results
were omitted), we can see significant improvements in security with
JEEP-C, up to 9\% for 0.1 bpnzAC. Finally, compared to the state-of-the-art
SVP without side information, we can see security gains of up to 27\%.

\subsection{ALASKA2}

\label{subsec:results_ALASKA}

Results for the ALASKA2 dataset are fairly similar to BOSSBase. We
can still see better security of SI-JMiPOD against DCTR. However,
SRNet is a more accurate detector, which in the end makes JEEP more
secure. We can observe that for payloads below 0.5 bpnzAC, JEEP is
a bit more detectable than SI-JMiPOD. We consider this difference
statistically insignificant since the detectability of both schemes
for these lower payloads is below 60\%. In other cases, JEEP offers
better security than the other two embedding schemes. This is especially
true for QF 95, where JEEP outperforms SI-JMiPOD by up to 3\% for
0.6 bpnzAC.

At QF 100, we again see improvement of JEEP-C over other side-informed
methods by up to 5.5\% in detectability at 0.1 bpnzAC. Compared to
non-informed SVP, the gains are as high as 29.5\%.

\section{Conclusions}

\label{sec:Conclusions}

This paper introduces a side-informed embedding scheme driven by a
statistical model of a decompressed image. For the first time, the
side information, in the form of JPEG compression rounding errors,
is used through covariance with the embedding changes. This allows
us to get rid of typical heuristics around the rounding errors. Since
we cannot compute the covariance from one given observation of a rounding
error, we estimate its value with the help of several simplifying
assumptions. This naturally creates an asymmetry of embedding changes,
which is captured by the Fisher information matrix.

Due to better steganalysis in the spatial domain, we use this covariance
in the model of a decompressed JPEG image. We then minimize the Kullback-Leibler
divergence between the cover and stego distributions to bound the
power of the likelihood ratio test. To be statistically significant,
we show that the side information can be and has to be considered
unavailable to a potential attacker.

We demonstrate through experiments that the proposed algorithm outperforms
other state-of-the-art side-informed algorithms. This is especially
true for images compressed with Quality Factor 100, where the Reverse
JPEG Compatibility Attack can be applied. By using constant pixel
variance, the gains in security are up to 9\%. In the future, we plan
to extend the methodology into color images, as well as different
types of side information.

The source code for JEEP will be made available from \href{https://janbutora.github.io/downloads/}{https://janbutora.github.io/downloads/}
upon acceptance of this paper.

\appendix{}
\begin{lem}
\label{lem:lemma1}If $\beta^{+}\geq2\beta^{-},u\geq0$, then

\[
3(\beta^{+}-\beta^{-})u\geq2(\beta^{+}+\beta^{-})u^{2}.
\]

Similarly for $\beta^{-}\geq2\beta^{+},u\leq0$.
\end{lem}
\begin{IEEEproof}
For $u=0$, the result holds. 

Let now $u\in(0,1/2]$ and $\beta^{+}\geq2\beta^{-}$. Let us assume
$3(\beta^{+}-\beta^{-})<2(\beta^{+}+\beta^{-})u$. We will show that
this leads to a contradiction.

It follows that 
\begin{eqnarray*}
\beta^{+}(3-2u) & < & \beta^{-}(2u+3)\\
\frac{\beta^{+}}{\beta^{-}} & < & \frac{2u+3}{3-2u}\\
 & = & -1+\frac{6}{3-2u}\\
 & < & 2.
\end{eqnarray*}
\end{IEEEproof}

\appendix{\begin{center}LIKELIHOOD-RATIO TEST\end{center}\label{appendix:LRT}}

For clarity, we omit the indices of variables in the following. The
test statistic of the LRT~\eqref{eq:LRT} for a single pixel is

\begin{eqnarray*}
\Lambda(e) & = & \frac{1}{2}\log\frac{\sigma^{2}}{\overline{\sigma}^{2}}+e^{2}\frac{\overline{\sigma}^{2}-\sigma^{2}}{2\sigma^{2}\overline{\sigma}^{2}}+e\frac{\mu}{\overline{\sigma}^{2}}-\frac{\mu^{2}}{2\overline{\sigma}^{2}}.
\end{eqnarray*}

It follows that the mean of the test statistic under null and alternative
hypotheses is

\begin{eqnarray*}
\mathbb{E}_{0}[\Lambda(e)] & = & \frac{1}{2}\log\frac{\sigma^{2}}{\overline{\sigma}^{2}}+\frac{\overline{\sigma}^{2}-\sigma^{2}}{2\overline{\sigma}^{2}}-\frac{\mu^{2}}{2\overline{\sigma}^{2}},\\
\mathbb{E}_{1}[\Lambda(e)] & = & \frac{1}{2}\log\frac{\sigma^{2}}{\overline{\sigma}^{2}}+\frac{\overline{\sigma}^{2}-\sigma^{2}}{2\sigma^{2}}+\mu^{2}\frac{2\sigma^{2}-\overline{\sigma}^{2}}{2\sigma^{2}\overline{\sigma}^{2}}.
\end{eqnarray*}
Next, let us remind the reader that for a random variable $X\sim\mathcal{N}(\mu,\sigma^{2})$,
the fourth moment is computed as $\mathbb{E}[X^{4}]=\mu^{4}+6\mu^{2}\sigma^{2}+3\sigma^{4}$,
and variance of $X^{2}$ is then given by $\mathrm{Var}(X^{2})=\mathbb{E}[X^{4}]-\sigma^{4}-\mu^{4}-2\mu^{2}\sigma^{2}=2\sigma^{4}+4\mu^{2}\sigma^{2}.$

The variances of a pixel test statistic are

\begin{eqnarray*}
\mathrm{Var}_{0}[\Lambda(e)] & = & \frac{\sigma^{2}}{\overline{\sigma}^{2}}(\overline{\sigma}^{2}-\sigma^{2}+\mu),\\
\mathrm{Var}_{1}[\Lambda(e)] & = & \frac{\overline{\sigma}^{2}+2\mu^{2}}{\sigma^{2}}(\overline{\sigma}^{2}-\sigma^{2})+\mu.
\end{eqnarray*}

The deflection coefficient $\delta$ and the variance effect $\varrho$
from~\eqref{eq:CLT} can now be expressed as
\begin{eqnarray*}
\delta & = & \frac{\sum_{i=1}^{N}\left(\mathbb{E}_{1}[\Lambda(e_{i})]-\mathbb{E}_{0}[\Lambda(e_{i})]\right)}{\sqrt{\sum_{i=1}^{N}\mathrm{Var}_{0}[\Lambda(e_{i})]}}\\
 & = & \frac{\sum_{i=1}^{N}\frac{\left(\overline{\sigma}_{i}^{2}-\sigma_{i}^{2}\right)^{2}+\mu_{i}^{2}(3\sigma_{i}^{2}-\overline{\sigma}_{i}^{2})}{2\sigma_{i}^{2}\overline{\sigma}_{i}^{2}}}{\sqrt{\sum_{i=1}^{N}\frac{\sigma_{i}^{2}}{\overline{\sigma}_{i}^{2}}(\overline{\sigma}_{i}^{2}-\sigma_{i}^{2}+\mu_{i})}},
\end{eqnarray*}

and 

\begin{eqnarray*}
\varrho & = & \frac{\sum_{i=1}^{N}\mathrm{Var}_{1}[\Lambda(e_{i})]}{\sum_{i=1}^{N}\mathrm{Var}_{0}[\Lambda(e_{i})]}\\
 & = & \frac{\sum_{i=1}^{N}\frac{\overline{\sigma}_{i}^{2}+2\mu_{i}^{2}}{\sigma_{i}^{2}}(\overline{\sigma}_{i}^{2}-\sigma_{i}^{2})+\mu_{i}}{\sum_{i=1}^{N}\frac{\sigma_{i}^{2}}{\overline{\sigma}_{i}^{2}}(\overline{\sigma}_{i}^{2}-\sigma_{i}^{2}+\mu_{i})}.
\end{eqnarray*}

We can see that even for realistic Eve ($\boldsymbol{\mu}=\mathbf{0})$,
expressions for $\delta$ and $\varrho$ are not possible to write
down in a simple form.

\appendix{\begin{center}TAYLOR EXPANSION OF KL DIVERGENCE\end{center}\label{appendix:taylor_expansion_KLD}}

Denote $D(\boldsymbol{\beta})=D(C||S)$ the $8\times8$ block KL divergence~\eqref{eq:KLD},
where $\boldsymbol{\beta}$ is a vector of all 128 change rates $\beta_{kl}^{+},\beta_{kl}^{-},\,0\leq k,l\leq7$.
Using the independence of embedding changes (S1), the derivatives
of stego mean and variance are

\begin{eqnarray*}
\frac{\partial\overline{\sigma}_{ij}^{2}}{\partial\beta_{kl}^{\pm}} & = & \left(f_{kl}^{ij}\right)^{2}q_{kl}^{2}((1\mp2u_{kl})^{2}\mp2(\beta_{kl}^{+}-\beta_{kl}^{-})),\\
\frac{\partial^{2}\overline{\sigma}_{ij}^{2}}{\partial^{2}\beta_{kl}^{\pm}} & = & -2\left(f_{kl}^{ij}\right)^{2}q_{kl}^{2},\\
\frac{\partial^{2}\overline{\sigma}_{ij}^{2}}{\partial\beta_{kl}^{+}\partial\beta_{kl}^{-}} & = & 2\left(f_{kl}^{ij}\right)^{2}q_{kl}^{2},\\
\frac{\partial\mu_{ij}}{\partial\beta_{kl}^{\pm}} & = & \pm f_{kl}^{ij}q_{kl}.
\end{eqnarray*}
Second-order Taylor expansion around zero dictates $D(\boldsymbol{\beta})=D(\mathbf{0})+\boldsymbol{\beta}\nabla D(\mathbf{0})+\frac{1}{2}\boldsymbol{\beta}\nabla^{2}D(0)\boldsymbol{\beta}^{T}.$

For zero-change rates, the stego distribution is equal to the cover
distribution, therefore $D(\mathbf{0})=0$. The derivatives of the
KL divergence w.r.t. change rates are

\begin{eqnarray*}
\frac{\partial}{\partial\beta_{kl}^{\pm}}D(\boldsymbol{\beta}) & = & \frac{\partial}{\partial\beta_{kl}^{\pm}}\frac{1}{2}\sum_{i,j=0}^{7}\log\frac{\overline{\sigma}_{ij}^{2}}{\sigma_{ij}^{2}}+\frac{\mu_{ij}^{2}+\sigma_{ij}^{2}-\overline{\sigma}_{ij}^{2}}{\overline{\sigma}_{ij}^{2}}\\
 & = & \frac{1}{2}\sum_{i,j=0}^{7}\left(\frac{\partial\overline{\sigma}_{ij}^{2}/\partial\beta_{kl}^{\pm}}{\overline{\sigma}_{ij}^{2}}-\frac{\sigma_{ij}^{2}\partial\overline{\sigma}_{ij}^{2}/\partial\beta_{kl}^{\pm}}{\overline{\sigma}_{ij}^{4}}\right.\\
 &  & \left.+\frac{2\mu_{ij}\overline{\sigma}_{ij}^{2}\partial\mu_{ij}/\partial\beta_{kl}^{\pm}-\mu_{ij}^{2}\partial\overline{\sigma}_{ij}^{2}/\partial\beta_{kl}^{\pm}}{\overline{\sigma}_{ij}^{4}}\right).
\end{eqnarray*}

Then, since $\overline{\sigma}_{ij}^{2}|_{\boldsymbol{\beta}=\mathbf{0}}=\sigma_{ij}^{2}$
we get $\nabla D(\mathbf{0})=\mathbf{0}$. We already see that the
only remaining term in the Taylor expansion is the Fisher information
matrix. We will show only the computation of $I_{kl}^{+}$ since $I_{kl}^{-}$
and $I_{kl}^{\pm}$ follow the same logic. 

Denoting $\partial\overline{\sigma}_{ij}^{2}=\partial\overline{\sigma}_{ij}^{2}/\partial\beta_{kl}^{+}$,
$\partial^{2}\overline{\sigma}_{ij}^{2}=\partial^{2}\overline{\sigma}_{ij}^{2}/\partial^{2}\beta_{kl}^{+}$
and $\partial\mu_{ij}=\partial\mu_{ij}/\partial\beta_{kl}^{+}$, the
second derivative of KL divergence is
\begin{eqnarray*}
\frac{\partial^{2}}{\partial^{2}\beta_{kl}^{+}}D(\boldsymbol{\beta}) & = & \frac{1}{2}\sum_{i,j}\left(\frac{\overline{\sigma}_{ij}^{2}\partial^{2}\overline{\sigma}_{ij}^{2}-(\partial\overline{\sigma}_{ij}^{2})^{2}}{\overline{\sigma}_{ij}^{4}}\right.\\
 &  & +\frac{2\overline{\sigma}_{ij}^{2}(\partial\mu_{ij})^{2}-\mu_{ij}^{2}\partial^{2}\overline{\sigma}_{ij}^{2}}{\overline{\sigma}_{ij}^{4}}\\
 &  & -\frac{\sigma_{ij}^{2}\overline{\sigma}_{ij}^{4}\partial^{2}\overline{\sigma}_{ij}^{2}-2\overline{\sigma}_{ij}^{2}\sigma_{ij}^{2}(\partial\overline{\sigma}_{ij}^{2})^{2}}{\overline{\sigma}_{ij}^{8}}\\
 &  & \left.-\frac{2\overline{\sigma}_{ij}^{2}\partial\overline{\sigma}_{ij}^{2}(2\mu_{ij}\overline{\sigma}_{ij}^{2}\partial\mu_{ij}-\mu_{ij}^{2}\partial\overline{\sigma}_{ij}^{2})}{\overline{\sigma}_{ij}^{8}}\right).
\end{eqnarray*}

Finally

\begin{eqnarray*}
I_{kl}^{+} & = & \frac{\partial^{2}}{\partial^{2}\beta_{kl}^{+}}D(\mathbf{0})\\
 & = & \frac{1}{2}\sum_{i,j=0}^{7}\frac{(\partial\overline{\sigma}_{ij}^{2}|_{\boldsymbol{\beta}=\mathbf{0}})^{2}}{\sigma_{ij}^{4}}+2\frac{(\partial\mu_{ij})^{2}}{\sigma_{ij}^{2}}\\
 & = & \frac{1}{2}\sum_{i,j=0}^{7}\frac{\left(f_{kl}^{ij}\right)^{4}q_{kl}^{4}(1-2u_{kl})^{4}}{\sigma_{ij}^{4}}+2\frac{\left(f_{kl}^{ij}\right)^{2}q_{kl}^{2}}{\sigma_{ij}^{2}}.
\end{eqnarray*}

\bibliographystyle{plain}
\bibliography{stegobooks}

\end{document}